\documentclass[a4paper,notitlepage,11pt]{article}

\usepackage[T1]{fontenc}   
\usepackage{graphicx} 
\usepackage[lmargin=1.7cm, rmargin=1.7cm,bottom=1.7cm,top=1.7cm]{geometry} 
\usepackage{amsmath}
\usepackage{amsthm} 
\usepackage{amssymb}  
\usepackage{url}
\usepackage{dsfont}
\usepackage{enumerate}
\usepackage{multicol}
\usepackage{array}
\usepackage{color}
\usepackage{caption}
\usepackage[rgb,usenames,dvipsnames]{xcolor}
\usepackage{cite}
\usepackage{hyperref}
\usepackage[rightcaption]{sidecap}  
\usepackage{algorithm} 
\usepackage{algorithmic} 
\usepackage{psfrag}
\usepackage{pgfplots}
\usepackage{tikz}
\usetikzlibrary{matrix,calc,decorations.pathreplacing,arrows,decorations.markings,external}
\newcommand*\xbar[1]{%
  \hbox{%
    \vbox{%
      \hrule height 0.5pt % The actual bar
      \kern0.5ex%         % Distance between bar and symbol
      \hbox{%
        \kern-0.1em%      % Shortening on the left side
        \ensuremath{#1}%
        \kern-0.1em%      % Shortening on the right side
      }%
    }%
  }%
}

\newcommand{\ZZ}[1]{\bigl\{\begin{smallmatrix} 0 \\ 0 \end{smallmatrix}\bigr\}_{#1}}
\newcommand{\OO}[1]{\bigl\{\begin{smallmatrix} 1 \\ 1 \end{smallmatrix}\bigr\}_{#1}}
\newcommand{\ZO}[1]{\bigl\{\begin{smallmatrix} 0 \\ 1 \end{smallmatrix}\bigr\}_{#1}}

\newcommand{\pattern}[3]{\bigl\{\begin{smallmatrix} #2 \\ #3 \end{smallmatrix}\bigr\}_{#1}}
\newcommand{\myhat}[1]{\hat{#1}}

\newtheorem{thm}{Theorem}

\newtheorem{prop}{Proposition}
\newtheorem{conj}{Conjecture}

\newtheorem{lem}{Lemma}

\theoremstyle{definition}
\newtheorem{defi}{Definition} 
\newtheorem{constr}{Construction}

\theoremstyle{remark}
\newtheorem{rem}{Remark}

\newtheoremstyle{defbis}% name of the style to be used
  {\topsep}% measure of space to leave above the theorem. E.g.: 3pt
  {\topsep}% measure of space to leave below the theorem. E.g.: 3pt
  {\normalfont}% name of font to use in the body of the theorem
  {0pt}% measure of space to indent
  {\bfseries}% name of head font
  {.}% punctuation between head and body
  {5pt plus 1pt minus 1pt}% space after theorem head; " " = normal interword space
  {\thmname{#1}\thmnumber{ \boldmath$#2^*$\unboldmath}\thmnote{ (#3)}}% Manually specify head
\theoremstyle{defbis}
\newtheorem{myconstr}{Construction} 

\newcommand{\mygrey}{black!70}

\begin{document}

\title { 
A complexity analysis of Policy Iteration \\ {\Large through combinatorial matrices arising from Unique Sink Orientations}%
\thanks{This work was supported by an ARC grant from the French Community of Belgium and by the IAP network 'Dysco' funded by the office of the Prime Minister of Belgium. The scientific responsiblity rests with the authors.}% 
}

\author{ 
Bal\'azs Gerencs\'er, 
Romain Hollanders, 
Jean-Charles Delvenne 
and Rapha\"el M. Jungers%
\thanks{The authors are with Department of Mathematical Engineering and ICTEAM at UCLouvain, 
4, avenue G. Lemaitre, B-1348 Louvain-la-Neuve, 
Belgium. J.-C. D. is with CORE and NAXYS. R. M. J. is an F.R.S./FNRS Research Associate. E-mail: \texttt{firstname.lastname@uclouvain.be}}%
}
%\date{\emph{April 2014}}
\date{}
\maketitle

\setlength{\parindent}{0pt}
\setlength{\parskip}{1ex plus 0.5ex minus 0.2ex}

\begin{abstract}
	Unique Sink Orientations (USOs) are an appealing abstraction of several major optimization problems of applied mathematics such as for instance Linear Programming (LP), Markov Decision Processes (MDPs) or 2-player Turn Based Stochastic Games (2TBSGs). A polynomial time algorithm to find the sink of a USO would translate into a strongly polynomial time algorithm to solve the aforementioned problems---a major quest for all three cases. 
%	In the case of an acyclic USO of a cube, a situation that captures both MDPs and 2TBSGs, one can apply 
	In addition, we may translate MDPs and 2TBSGs into the problem of finding the sink of an acyclic USO of a cube, which can be done using
	the well-known Policy Iteration algorithm (PI). The study of its complexity is the object of this work. Despite its exponential worst case complexity, the principle of PI is a powerful source of inspiration for other methods. 
	
	As our first contribution, we disprove Hansen and Zwick's conjecture claiming that the number of steps of PI should follow the Fibonacci sequence in the worst case. Our analysis relies on a new combinatorial formulation of the problem---the so-called Order-Regularity formulation (OR). Then, for our second contribution, we (exponentially) improve the $\Omega(1.4142^n)$ lower bound on the number of steps of PI from Schurr and Szab\'o in the case of the OR formulation and obtain an $\Omega(1.4269^n)$ bound.
\end{abstract}

\section{Introduction}

\textbf{Three problems.} Optimizing a linear function under a set of linear constraints is one of the most successful problems in engineering well known as \emph{Linear Programming} (LP). Decision making in a stochastic environment is conveniently modeled using \emph{Markov Decision Processes} (MDPs). Finding an optimal strategy for the Backgammon board game can be modeled as a \emph{2-player Turn-Based Stochastic Game} (2TBSG). 
%These three tools all have a community of their own. Yet, they have an important common point: 
These three vastly studied problems share an important common point: 
%they can all be translated into a single mathematical formulation called \emph{Unique Sink Orientations} (USOs) and solved using any USO algorithm to find the \emph{sink}.
they can all be seen as special families of instances of the problem of finding the \emph{sink} of a \emph{Unique Sink Orientation} (USO).

\textbf{Unique Sink Orientations, a rich structure.} Introduced by Szab\'o and Welzl~\cite{szabo2001}, Unique Sink Orientations are an appealing extension for many frameworks including LP, MDPs, 2TBSGs~\cite{hansen2014,adler1974,condon1992,ludwig1995}, but also \emph{Linear Complementarity Problems}~\cite{gartner2005,cottle2009,stickney1978} or the problem of finding the \emph{smallest enclosing ball} to a set of points~\cite{szabo2001,gartner2001} for instance. Algorithms that find the sink of USOs can also be used to solve any of the aforementioned problems. 
In general, a USO of a polytope is an orientation of its edges such that any face at any dimension has a unique sink (that is, a unique vertex with only incoming links). In particular, this implies that the whole polytope allows a unique sink.
Polytopes arising from LP naturally exhibit a USO structure, where edges are directed towards better objective values (we here exclude degeneracies, such as when two vertices have the same objective value). In this case, the polytopes have the additional property of being \emph{acyclic}. We then talk about Acyclic USOs, or AUSOs, and the vertices of the polytope form a partial order. The global sink corresponds to the optimal solution of the corresponding LP.
MDPs and 2TBSGs exhibit an even more special structure as they can be represented as the AUSO of a \emph{hypercube} (or Cube AUSO). Cube (A)USOs are convenient for algorithmic purposes because any vertex can be queried at any time, unlike Simplex-like methods that do only allow queries from neighbor to neighbor.
Interestingly, G\"artner and Schurr showed that an LP can always be formulated as a Cube USO whose solution translates either to the solution of the LP, or into a certificate of unboundedness or infeasibility~\cite{gartner2006}. However they do not mention whether the corresponding USO is acyclic or not.

\textbf{Which algorithms to find the sink?} A major goal in the study of (A)USOs is to find a polynomial time algorithm to find the sink. More precisely, such an algorithm should make no more than a polynomial number of vertex evaluations in the dimension or the number of facets of the (A)USO. By ``vertex evaluation'', we mean a request of the orientations of the edges adjacent to the vertex. In the case of LP, answering this question for Cube USOs would imply the first strongly polynomial time algorithm, a long lasting quest since the first weakly polynomial time algorithms were found about 30 years ago (namely \emph{interior point} methods~\cite{karmarkar1984} and \emph{ellipsoids} methods~\cite{khachiyan1980}). Regarding MDPs and 2TBSGs, it would be enough to find a polynomial time algorithm for Cube AUSOs to obtain the same consequence. Note that an MDP can always be formulated as an LP so it is not surprising that the problem at hand is simpler. It is more surprising for 2TBSGs though, as today no polynomial time algorithms are known to solve them in general~\cite{hansen2013}.

Most algorithms to find the sink of a Cube (A)USO can be categorized along two axes: 
\begin{itemize}
	\item they can be \emph{deterministic} or \emph{randomized} when choosing the next vertex to query;
	\item at two successive steps, they can perform \emph{local hops} (like the Simplex algorithm, from neighbor to neighbor) or large \emph{jumps} in the cube.
\end{itemize} 
A successful example of a jumping deterministic algorithm is the so-called \emph{Fibonacci Seesaw} that applies to both USOs and AUSOs and is guaranteed to converge in at most around $O(1.61^n)$ steps, $n$ being the dimension of the cube~\cite{szabo2001}. This is today the best known upper bound for deterministic (A)USO algorithms. On the other hand, \emph{Random-Facet} for AUSOs is a local-hop randomized algorithm that is currently the only known method to solve AUSOs in sub-exponential time, namely in at most $e^{O(\sqrt{n})}$ steps~\cite{gartner2002}. However, this bound was shown to be essentially tight~\cite{matousek1994}. Apart from these two, many other methods have received quite a lot of attention, such as the \emph{Product} Algorithm, \emph{Random-Edge} or \emph{Random-Jump} to mention only a few~\cite{szabo2001,hansen2014,gartner1998,morris2002,Mansour1999}. 

\textbf{Policy Iteration, inspiring dynamics.} An interesting competitor to the Fibonacci Seesaw---and the focus of this paper---is the \emph{Policy Iteration} algorithm (PI), a deterministic jumping algorithm specialized for AUSOs. First introduced by Howard for MDPs~\cite{Howard1960}, it was later adapted for \emph{Parity Games}---a variant of 2TBSGs~\cite{rao1973}---and AUSOs~\cite{schurr2005}. 
%PI is also referred to as Strategy Iteration in the Parity Games framework or as Bottom-Antipodal or Jump in the AUSO framework.
PI's update rule is based on the following property: let $v$ be a vertex of an AUSO and let $u$ be any vertex in the sub-cube rooted in $v$ and spanned by the out-links of $v$, then there is always a directed path from $v$ to $u$. Intuitively, because of the acyclicity of the orientation, this means that any vertex $u$ is ``closer'' to the sink than $v$. From $v$, choosing any vertex $u \neq v$ in the sub-cube as the next iterate would therefore provide an update rule that is guaranteed to converge to the global sink. PI's choice is to jump to the vertex $u$ that is \emph{antipodal} to $v$ in the sub-cube and iterate from there. Therefore, it is sometimes referred to as \emph{Bottom-Antipodal} or \emph{Jump}. Note that Random-Jump mentioned above is based on the same principle except that it chooses any vertex of the sub-cube with uniform probability~\cite{Mansour1999}.

\textbf{Complexity issues.} Today's best upper bound on the complexity of PI for general MDPs and 2TBSGs is due to Hollanders et al. with a $(2+o(1)) \cdot 2^n/n$ bound on the number of steps~\cite{hollanders2014b}, improving on the $O\big(2^n/n\big)$ bound from Mansour and Singh~\cite{Mansour1999}. 
%In some cases, it was shown to run in polynomial time such as for \emph{discounted} MDPs and 2TBSGs with a fixed discount factor~\cite{Tseng1990,Ye2011} or \emph{deterministic} MDPs and 2TBSGs~\cite{post2013} (the bounds in these results were later improved in \cite{hansen2013} and \cite{scherrer2013}). 
In some cases such as \emph{discounted} MDPs and 2TBSGs with a fixed discount factor, PI was shown to run in strongly polynomial time~\cite{Tseng1990,Ye2011}. A close variant of PI (namely a Simplex-like version) was also shown to run in strongly polynomial time for \emph{deterministic} MDPs and 2TBSGs~\cite{post2013}, leaving open the question for the usual PI. (The polynomial bounds in the two results were later improved in \cite{hansen2013} and \cite{scherrer2013}.) 
Regarding lower bounds, Friedmann exhibited a construction where PI requires at least $\Omega(\sqrt[9]{2}^n)$ steps to converge for Parity Games~\cite{friedmann2009}. This bound was then adapted to MDPs by Fearnley and Hollanders et al. as an $\Omega(\sqrt[7]{2}^n)$ lower bound~\cite{Fearnley2010,hollanders2012}. (Fearnley translated Friedmann's construction to total- and average-cost MDPs and Hollanders et al. to discounted-cost MDPs.) For AUSOs, Schurr and Szab\'o provided an $\Omega(\sqrt{2}^n)$ bound~\cite{schurr2005}.
Nevertheless, PI remains a very efficient algorithm in practice.
Unfortunately, no convergence guarantees for cyclic USOs exist for PI.

\textbf{A new formulation, towards new bounds.} In this work, we investigate a new relaxation on the maximal number of steps taken by PI in a Cube AUSO known as the \emph{Order-Regularity} problem (OR). First introduced by Hansen~\cite{Hansen2012}, the idea of this formulation is the following. Suppose that PI explores a sequence of vertices $v_1, v_2, ..., v_m$ in an AUSO. Here we represent vertices of the $n$-dimensional cube by an $n$-dimensional binary tuple. In the OR formulation, we record all the vertices into an $m \times n$ binary matrix so that each row corresponds to an iteration. We then translate the AUSO property into a combinatorial condition on this matrix, as defined in Section~\ref{sec:preliminaries}, Definition~\ref{def:OR}.

Using the OR formulation, Hansen and Zwick performed an exhaustive search on all OR matrices with up to $6$ columns and reported the maximum number of rows (that is, iterations for PI) each time: $2, 3, 5, 8, 13, 21$. Based on these empirical observation, they conjectured that the maximum number of steps of PI should follow the \emph{Fibonacci sequence}~\cite{Hansen2012}. Confirming this conjecture for $n=7$ has been claimed to be a hard computational challenge. It was introduced as January 2014's \emph{IBM Ponder This} Challenge. Proving the conjecture in general would provide an $O(1.618^n)$ upper bound on the number of iterations of PI, a quasi-identical bound as that for Fibonacci Seesaw.
Regarding lower bounds, no better bound than the one from Schurr and Szab\'o for the AUSO framework was known prior to our work.

\textbf{Results.} In this paper, our first contribution is to disprove Hansen and Zwick's conjecture by performing an exhaustive search for $n=7$. We obtained a maximum number of rows that is lower than the expected Fibonacci number, which suggests that matching or even improving the $O(1.61^n)$ bound of Fibonacci Seesaw may be possible.
Our second contribution is to (exponentially) improve Schurr and Szab\'o's $\Omega(1.4142^n)$ lower bound to $\Omega(1.4269^n)$, yet only in the framework of OR matrices.
The key ideas behind our results both rely on the construction of large matrices satisfying OR-like conditions, which required substantial computational refinements.

\textbf{Structure.} The paper is organized as follows. In Section~\ref{sec:preliminaries} we formulate the OR condition together with some key elements for its analysis. In Section~\ref{sec:fibonacci-conjecture}, we discuss Hansen and Zwick's conjecture and formulate our first result. Section~\ref{sec:lower-bound} establishes our new lower bound step by step on the number of rows of OR matrices, starting from Schurr and Szab\'o's construction. Then, since our results heavily rely on our ability to build large matrices, we describe in Section~\ref{sec:large-matrices} the different ideas that we combined in order to speed up the existing computational methods. 
Finally, we conclude with some perspectives in Section~\ref{sec:conclusions}.

\section{Problem setting and preliminaries} \label{sec:preliminaries}

\begin{defi}[Order-Regularity~\cite{Hansen2012}] \label{def:OR}
We say that $A \in \{0,1\}^{m \times n}$ is Order-Regular (OR) whenever for every pair of rows $i,j$ of $A$ with $1 \leq i < j \leq m$, there exists a column $k$ such that 
\begin{align} \label{eq:OR}
	A_{i,k} \neq A_{i+1,k} = A_{j,k} = A_{j+1,k}.
\end{align}
We may have $j + 1 = m + 1$. In that case, we use the convention that $A_{m+1,k} = A_{m,k}$.
%In case $j+1 > m$, we use the convention $A_{m+1,k} = A_{m,k}$.
Furthermore, the last two rows (labeled $m-1$ and $m$) are required to be distinct.
\end{defi}

With other words, for all pairs $(i,j)$, there exists a column $k$ in $A$ such that at the entries $i, i+1, j, j+1$ in this column we see either $0, 1, 1, 1$ or $1, 0, 0, 0$. Another possible reading of the OR condition in terms of Policy Iteration is as follows: at any iteration, some changes are made to the entries of the current vertex.
It must always be assumed in future iterations that at least one of these changes was ``right''.
%is that when some changes are made to a row when stepping to the next, then not all of them can be wrong at once and part of these changes must still be there, and maintained, at any later step. 

Note that the OR condition is invariant under permutation or negation of its columns. By \emph{negating} a column, we mean changing all 0 entries to 1 and vice versa. Also observe that each pair $(i,j)$ can be considered as a constraint to be satisfied by some column of the matrix. We will say that $A$ \emph{satisfies} a constraint $(i,j)$ if it satisfies condition~\eqref{eq:OR} for some column $k$. 
%If a matrix fails to satisfy some constraints, adding columns can only help to satisfy these missing constraints.

Our aim is to find bounds on the number of rows of Order-Regular matrices.

\begin{defi}[Constraint space]
We introduce the \emph{constraint space} as a visualization tool to relate a binary matrix with the Order-Regularity condition. To any pair $(i,j)$ for which condition \eqref{eq:OR} is required (i.e. for all $i,j$ such that $1 \leq i < j \leq m$), we associate a unit square of the Cartesian grid centered at coordinate $(i,j)$. 
Whenever we want to emphasize which part of the matrix satisfies which part of the constraint space, we use a matching coloring on some subset of entries of the matrix and the corresponding squares of the constraint space.
%Let $C:\{1,\cdots,m\} \times \{1,\cdots,n\} \rightarrow \mathbb{N}$ be a coloring of the entries of $A$; if for some rows $i,j$, there exists a column $k$ such that condition \eqref{eq:OR} is verified and such that $C(i,k) = C(j,k) \triangleq c$, then we associate color $c$ to the square corresponding to $(i,j)$ in the constraint space. It happens that several colors can be used for a square; in that case any of the allowed colors can be chosen, whichever is the most convenient. We define the color white (or 0) as the \emph{undefined} color that can always be chosen.
\end{defi}

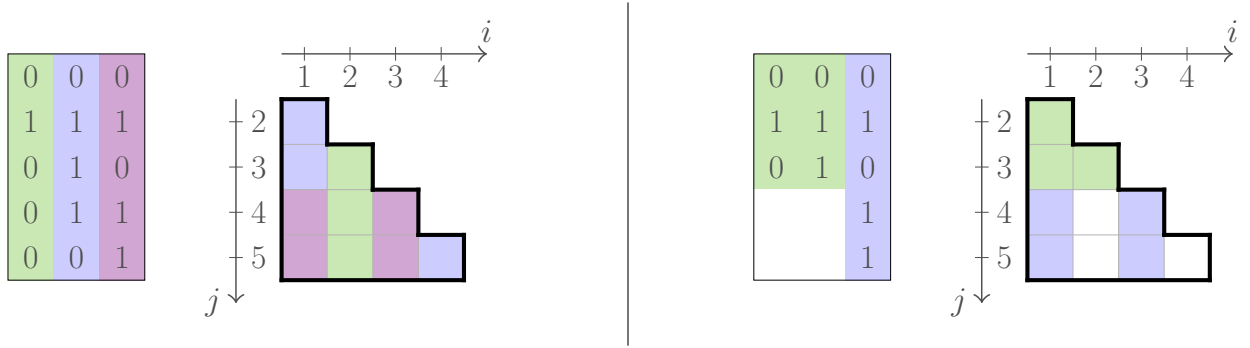
\begin{figure}[h!]
\begin{center}
\begin{tabular}{cc|cc}
\scalebox{.6}{
\pgfimage{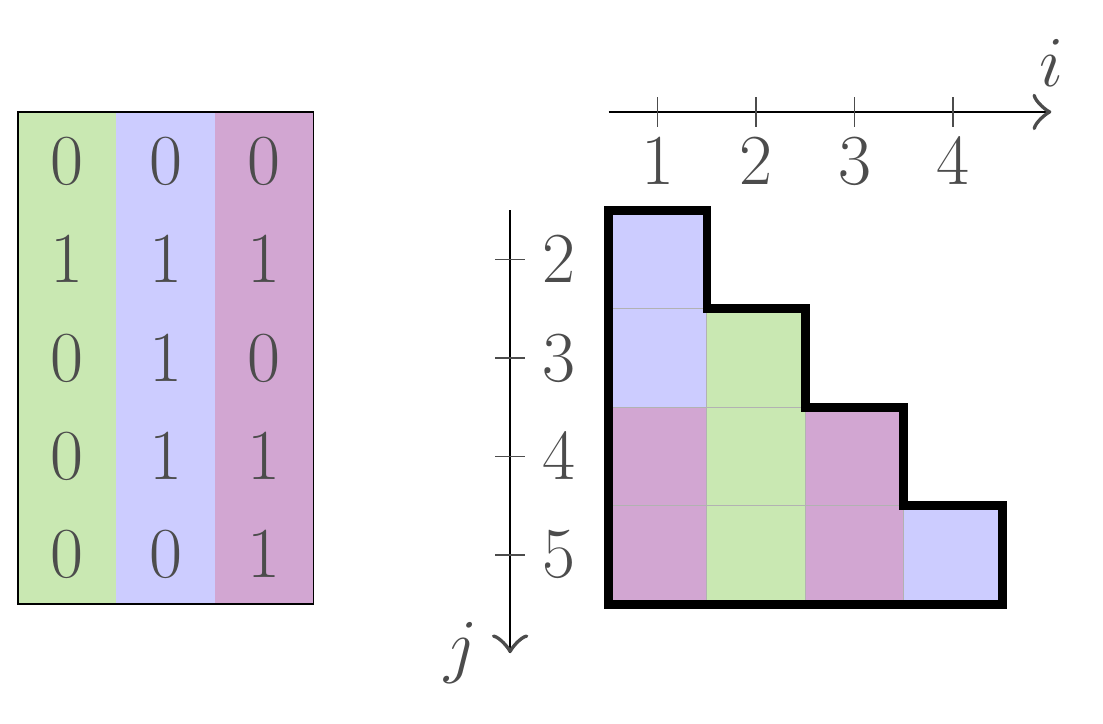}
}
&
\scalebox{.6}{
\tikzsetnextfilename{foo1}
\begin{tikzpicture}\path[color=white] (0,0) -- (1,0);\end{tikzpicture}
}
&
\scalebox{.6}{
\tikzsetnextfilename{foo2}
\begin{tikzpicture}\path[color=white] (0,0) -- (1,0);\end{tikzpicture}
}
&
\scalebox{.6}{
\pgfimage{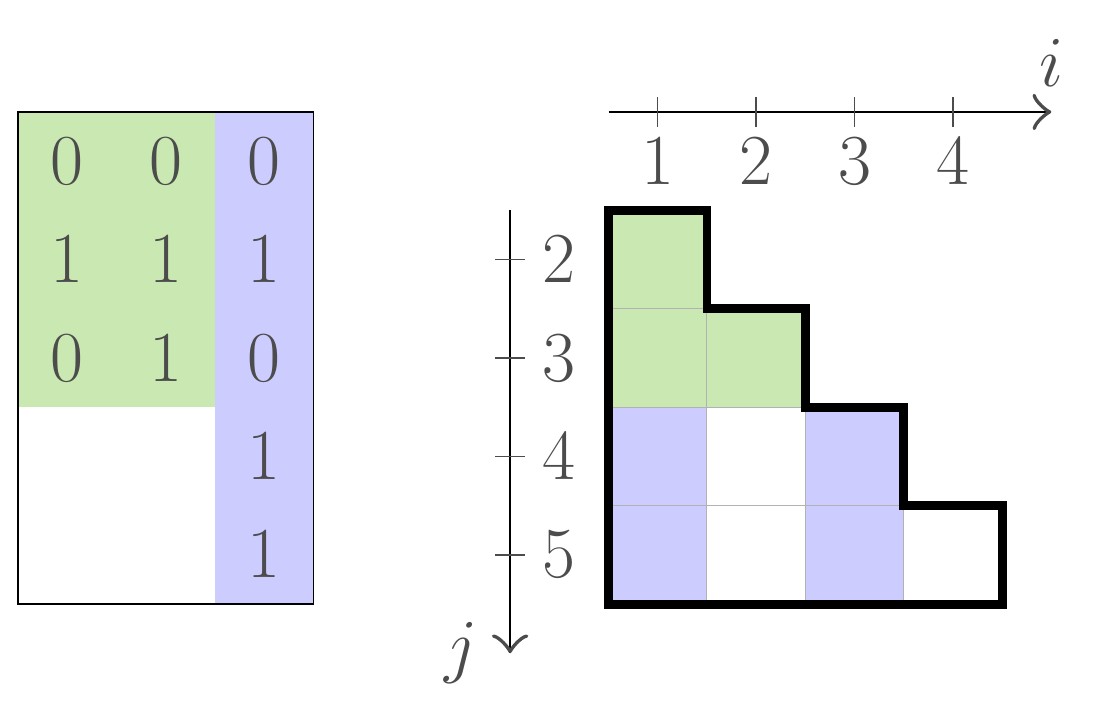}
}
\end{tabular}
\end{center}
\vspace{-.5cm}
\caption{\small (Left) Each column of the matrix and the constraints it satisfies are associated with a color. (Right) The contribution of the red and blue blocks are indicated while the unfilled part of the matrix is disregarded.}
\label{fig:constraint-space-example}
\end{figure}

The constraint space can be used in several ways. When considering a matrix that is not Order-Regular, it allows to visualize which constraints $(i,j)$ are satisfied by the matrix and which ones are not, and possibly detect patterns. It also allows to visualize how each column of a matrix (or even any given part of it) contributes in achieving Order-Regularity as illustrated by Figure~\ref{fig:constraint-space-example}.

Most binary vectors encountered in our constructions 
are repetitions of simple patterns, for which we introduce a compact notation.
%are entirely defined by their two first entries that are then repeated the appropriate number of times. A similar observation goes with matrices that are entirely defined by two smaller matrices that are repeated thereafter. 
%Therefore, we introduce a compact notation for these patterns.

\begin{defi}[Patterns]
	A \emph{pattern} $\pattern{M}{A}{B}$ is a matrix composed of $M$ copies of $A$ or $B$ put below each other in alternance, starting from $A$. Here $M$ is called the size of the pattern. The matrices $A$ and $B$ are assumed to have the same number of columns. For example, $\ZO{5} = \bigl[\begin{matrix} 0 & 1 & 0 & 1 & 0 \end{matrix}\bigr]^T$. We sometimes omit the size of pattern if clear from the context.
\end{defi}

The following operation is also frequently used in our constructions.

%\begin{defi}[Gluing]
%	Let $A'$ and $A''$ be two binary matrices with the same number of columns $n$ and let $\widetilde{A}''$ be obtained from $A''$ by negating some of its columns such that the first row of $\widetilde{A}''$ is identical to the last row of $A'$. We call $A = \left[ \begin{smallmatrix} A' \\ \widetilde{A}'' \end{smallmatrix} \right]$ the \emph{gluing} of $A'$ and $A''$. 
%	We call \emph{$M$-gluing} of $A'$ and $A''$ the construction of a matrix $A$ composed of $M$ alternating copies of $A'$ and $\widetilde{A}''$ such that:
%	\begin{align*}
%		A= \left\{ \begin{matrix} A' \\ \widetilde{A}'' \end{matrix} \right\}_{\!M} 
%			\color{black!50}{= \begin{bmatrix} A' \\ \widetilde{A}'' \\ A' \\ \vdots \\ A' \end{bmatrix}}.
%	\end{align*}
%	Notice that the first row of $A'$ is also identical to the last row of $\widetilde{A}''$. If $A'' = A'$, we call $A$ the $M$-gluing of $A'$. The effect of gluing is illustrated in Figure~\ref{fig:gluing} with two Order-Regular matrices.
%%	it satisfies condition~\eqref{eq:OR} for every pairs $(i,j)$ such that $1 \leq i < j < m'$ or $m'+1 \leq i < j \leq m''$. The strict inequality $j < m'$ comes from the fact that the convention $A_{m',k} = A_{m'+1,k}$ for all $1 \leq k \leq n$ from Definition~\ref{def:OR} is not guaranteed because the last row of $A'$ is not necessarily identical to the first row of $A''$. If it is the case though, then Order-Regularity is preserved even when $j = m'$. We then say that the gluing is \emph{proper}. 
%\end{defi}

\begin{figure}[h!]
\begin{center}
\scalebox{.6}{
\pgfimage{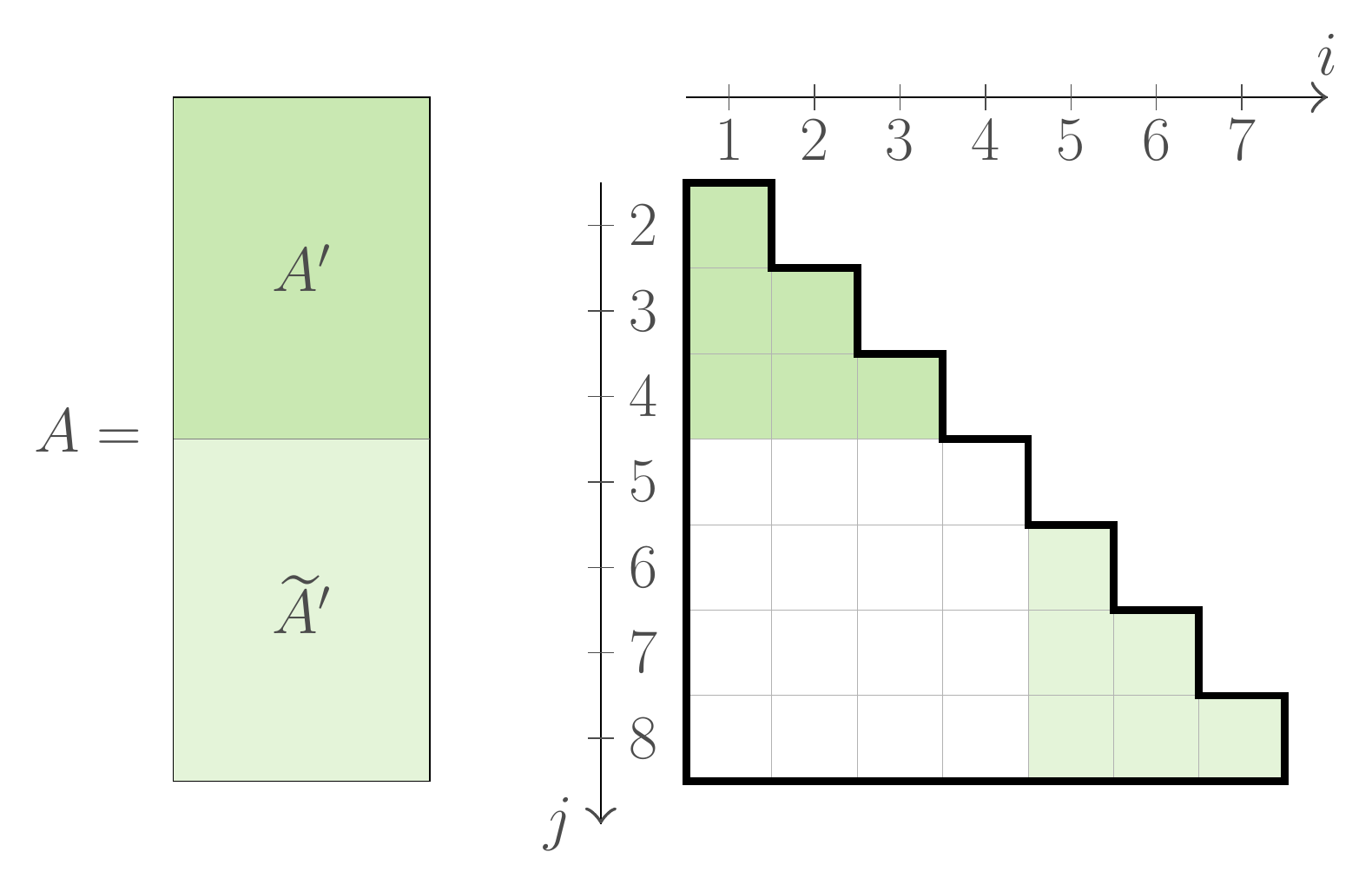}
}
\end{center}
\vspace{-.5cm}
\caption{\small An illustration of the effect of a 2-gluing with a 4-rows matrix $A$. If the last row of $A$ had been different from the first row of $\widetilde{A}$, then the row of $j=4$ of the constraint space would not have been completely colored.}
\label{fig:gluing}
\end{figure}

\begin{defi}[Gluing]
	Let $A$ be a binary matrix and let $\widetilde{A}$ be obtained from $A$ by negating some of its columns such that the first row of $\widetilde{A}$ is identical to the last row of $A$. 
%	We call $A' = \left[ \begin{smallmatrix} A \\ \widetilde{A} \end{smallmatrix} \right]$ the \emph{gluing} of $A$. 
	We call ``the \emph{$M$-gluing} of $A$'' the construction of a matrix $A'$ composed of $M$ alternating copies of $A$ and $\widetilde{A}$ such that:
	\begin{align*}
		A' = \left\{ \begin{matrix} A \\ \widetilde{A} \end{matrix} \right\}_{\!M} 
			= \begin{bmatrix} A \\ \widetilde{A} \\ \vdots \\ A \\ \widetilde{A} \\ A \end{bmatrix}.
	\end{align*}
	Notice that if $A$ is OR, then $\widetilde{A}$ is OR too. Furthermore, the first row of $A$ is also identical to the last row of $\widetilde{A}$. The effect of gluing is illustrated in Figure~\ref{fig:gluing} with $A$ an OR matrix.
%	it satisfies condition~\eqref{eq:OR} for every pairs $(i,j)$ such that $1 \leq i < j < m'$ or $m'+1 \leq i < j \leq m''$. The strict inequality $j < m'$ comes from the fact that the convention $A_{m',k} = A_{m'+1,k}$ for all $1 \leq k \leq n$ from Definition~\ref{def:OR} is not guaranteed because the last row of $A'$ is not necessarily identical to the first row of $A''$. If it is the case though, then Order-Regularity is preserved even when $j = m'$. We then say that the gluing is \emph{proper}. 
\end{defi}

The following straightforward lemma will be of key importance for our constructions. 

\begin{lem} \label{thm:effect-of-gluing}
	Let $A'$ be the $M$-gluing of an OR matrix $A$ with $m$ rows. Then any constraint $(i,j)$ such that $(s-1) \cdot m < i < j \leq s \cdot m$ for some integer $1 \leq s \leq M$ is satisfied by $A'$. 
\end{lem}

Notice that if the appropriate columns of $A \in \{0,1\}^{m \times n}$ had not been negated to obtain $\widetilde{A}$, then Lemma~\ref{thm:effect-of-gluing} would not have been guaranteed when $j = s \cdot m$ for any $1 \leq s < M$. This is because the convention $A_{s \cdot m,k}' = A_{s \cdot m+1,k}'$ from Definition~\ref{def:OR} must be ensured for all columns at the connection between the different $A$ and $\widetilde{A}$ blocks.

%\section{Former results}
%
%
%
%The results that we present in this paper react to some existing results and constructions from the literature. One is about disproving Hansen and Zwick's conjecture about the size of extremal Order-Regular matrices. The other is about improving Schurr and Szabó's lower bound on the complexity of Bottom-Antipodal to solve Acyclic Unique Sink Orientations, a lower bound that we translated in the framework of Order-Regular matrices. We now state these two former results.
%
%
%
%\subsection{\boldmath Extremal Order-Regular matrices up to $n = 6$\unboldmath}

\section{Refuting the Fibonacci Conjecture} \label{sec:fibonacci-conjecture}

In their works, Hansen and Zwick have performed an exhaustive search on every possible Order-Regular matrix with up to $n = 6$ columns and proposed the following conjecture that matches their observations perfectly.

\begin{conj}[Hansen and Zwick, 2012~\cite{Hansen2012}] \label{thm:hansen-zwick-conjecture}
	The maximum number of rows of an $n$-column Order-Regular matrix is given by $F_{n+2}$, the $(n+2)^{\text{nd}}$ Fibonacci number.
\end{conj}

Note that except for $n = 5$, the extremal matrix found (that is, the matrix with the most rows) was always unique, up to symmetry. In Table~\ref{fig:order-regular-examples}, we show these extremal matrices for $n = 3$ and $4$.

\begin{table}[h!]
  \centering
  {\footnotesize\tt
  \begin{tabular}{|ccc|}
  	\hline
    0&0&0\\
    1&1&1\\
    0&0&1\\
    0&1&1\\
    0&1&0\\
    \hline
  \end{tabular}
  \hspace{1cm}
  \begin{tabular}{|cccc|}
  	\hline
    0&0&0&0\\
    1&1&1&1\\
    0&0&0&1\\
    0&1&1&1\\
    0&0&1&0\\
    0&1&1&0\\
    0&1&0&0\\
    1&1&0&0\\
    \hline
  \end{tabular}}
  \caption{The unique extremal Order-Regular matrices for 3 and 4 columns.}
  \label{fig:order-regular-examples}
\end{table}

Whereas Conjecture~\ref{thm:hansen-zwick-conjecture} suggests that extremal OR matrices should have at most $O\big(\phi^n\big)$ rows, with $\phi = (\sqrt{5}+1)/2 \approx 1.618$ the golden ratio, the best proven upper bound on their number of rows is presently much higher.

\begin{prop} \label{thm:upper-bound-OR-matrices}
	Let $A \in \{0,1\}^{m \times n}$ be an Order-Regular matrix. Then $m \leq 2^{n-1}+1$.
	
	\begin{proof}
		Let us relax the Order-Regularity condition~\eqref{eq:OR} and require the following condition instead:
		\begin{align} \label{eq:relaxed-OR}
			A_{i,k} \neq A_{i+1,k} = A_{j,k},
		\end{align}
		that is, we no longer require that $A_{j,k} = A_{j+1,k}$. Let $A' \in \{0,1\}^{m' \times n}$ be an extremal matrix for this relaxed OR condition for a given number of columns $n$. Clearly $m'$ is at least as large as the number of rows of any OR matrix with $n$ columns since $A'$ is strictly less constrained. We now show that $m'$ is bounded from above by $2^{n-1}+1$ to obtain the result. Indeed, for any rows $i,j$, $1 \leq i < j \leq m'$, we can never have $A'_{i+1,k} \neq A'_{j,k}$ for all columns $k$. Therefore, $A'$ can never contain both a row and its negation except if one of the two is the first row. Hence it cannot have more than $2^{n-1}+1$ rows.
	\end{proof}
\end{prop}

Note that the above bound is optimal for the relaxation mentioned in the proof. However, the proof of this statement is beyond the scope of this paper.

In Section~\ref{sec:large-matrices}, we develop an algorithm to search for large OR matrices, with a special attention given to speed. Using this algorithm, we were able to perform an exhaustive search on all OR matrices with $n=7$ columns. The largest matrices we found had only 33 rows, hence the following Theorem.

\begin{thm} \label{thm:conjecture-fails}
	For $n = 7$, there exist no Order-Regular matrix with $34 = F_{n+2}$ rows and therefore Conjecture~\ref{thm:hansen-zwick-conjecture} fails.
\end{thm}

%Of course, a formal proof of Theorem~\ref{thm:conjecture-fails} would require a proof of correctness for our code. While we cannot provide such a proof, 
The proof of Theorem~\ref{thm:conjecture-fails} is given by the computer-aided exhaustive search for $n=7$.
Our implementation of the algorithm described in Section~\ref{sec:large-matrices} is available at \url{http://perso.uclouvain.be/romain.hollanders/docs/GoCode/ORsearch.zip}. 
%We also encourage anyone interested to implement their own variant of the algorithm and thereby confirm our result.

\section{New lower bounds on the size of Order-Regular matrices} \label{sec:lower-bound}

This section is organized as follows. First, we show a simple construction that allows to build OR matrices with $n$ columns and $\Omega(\sqrt{2}^n)$ rows. Then we detail our construction to beat this bound as follows:
\begin{enumerate}
	\item we introduce Strongly Order-Regular matrices, a refinement of OR matrices that we will need for our construction and improve on the $\sqrt{2}$ growth rate of the number of rows from the simple construction;
	\item we describe and prove the heart of our construction and obtain a first improvement over Schurr and Szab\'o's bound in the setting of OR matrices;
	\item we finally add an additional refinement to our construction that allows to improve our bound even a bit further to our final bound.
\end{enumerate}

\subsection{\boldmath A family of Order-Regular matrices with $\Omega(\sqrt{2}^n)$ rows\unboldmath}

The following construction provides Order-Regular matrices for arbitrarily large $n$ with $m=\Omega\big(\sqrt{2}^n\big)$. 

\begin{constr} \label{def:simple-construction}
	We recursively build a matrix $A^{(\ell)}$ as follows:
	\[ 
		A^{(\ell)} = 
			\begin{bmatrix} 
				\begin{array}{c|cc}
					&&\\
					\qquad A^{(\ell-1)} \qquad \color{white}{.}
					&
%					\begin{matrix}
%						0 & 0 \\ 1 & 1 \\ 0 & 0 \\ 1 & 1 \\ \vdots & \vdots \\ 0 & 0 \\ 1 & 1 
%					\end{matrix} \\
					\ZO{}
					&
					\ZO{} \\
					&&\\
					\hline
					&&\\
					\qquad \widetilde{A}^{(\ell-1)} \qquad \color{white}{.} 
					&
%					\begin{matrix} 
%						1 & 0 \\ 1 & 0 \\ 1 & 0 \\ 1 & 0 \\ \vdots & \vdots \\ 1 & 0 \\ 1 & 0 
%					\end{matrix}
					\OO{}
					&
					\ZZ{}\\
					&&
				\end{array}
			\end{bmatrix}
	\] 
	where $A^{(1)} = \left[ \begin{smallmatrix} 0 \\ 1 \end{smallmatrix} \right]$ and where $\widetilde{A}^{(\ell-1)}$ is obtained from $A^{(\ell-1)}$ by negating some of its columns such that the first row of $\widetilde{A}^{(\ell-1)}$ is identical to the last row of $A^{(\ell-1)}$. The size of all the patterns used in the construction is $2^{\ell-1}$ and the resulting matrix $A^{(\ell)}$ has $n_{\ell} = 2\ell-1$ columns and $m_{\ell} = 2^{\ell}$ rows.
%	It is easy to verify that $A^{(1)}$ is Order-Regular and therefore that every iterate of this recursive construction is Order-Regular as well. Furthermore, $A^{(\ell)}$ has $2^{\ell}$ rows, hence the $\Omega\big(\sqrt{2}^n\big)$ lower bound. 
	A matching construction is given in~\cite{schurr2005} in the framework of Acyclic Unique Sink Orientations.
\end{constr}

\begin{figure}[h!]
\begin{center}
\scalebox{.6}{
\pgfimage{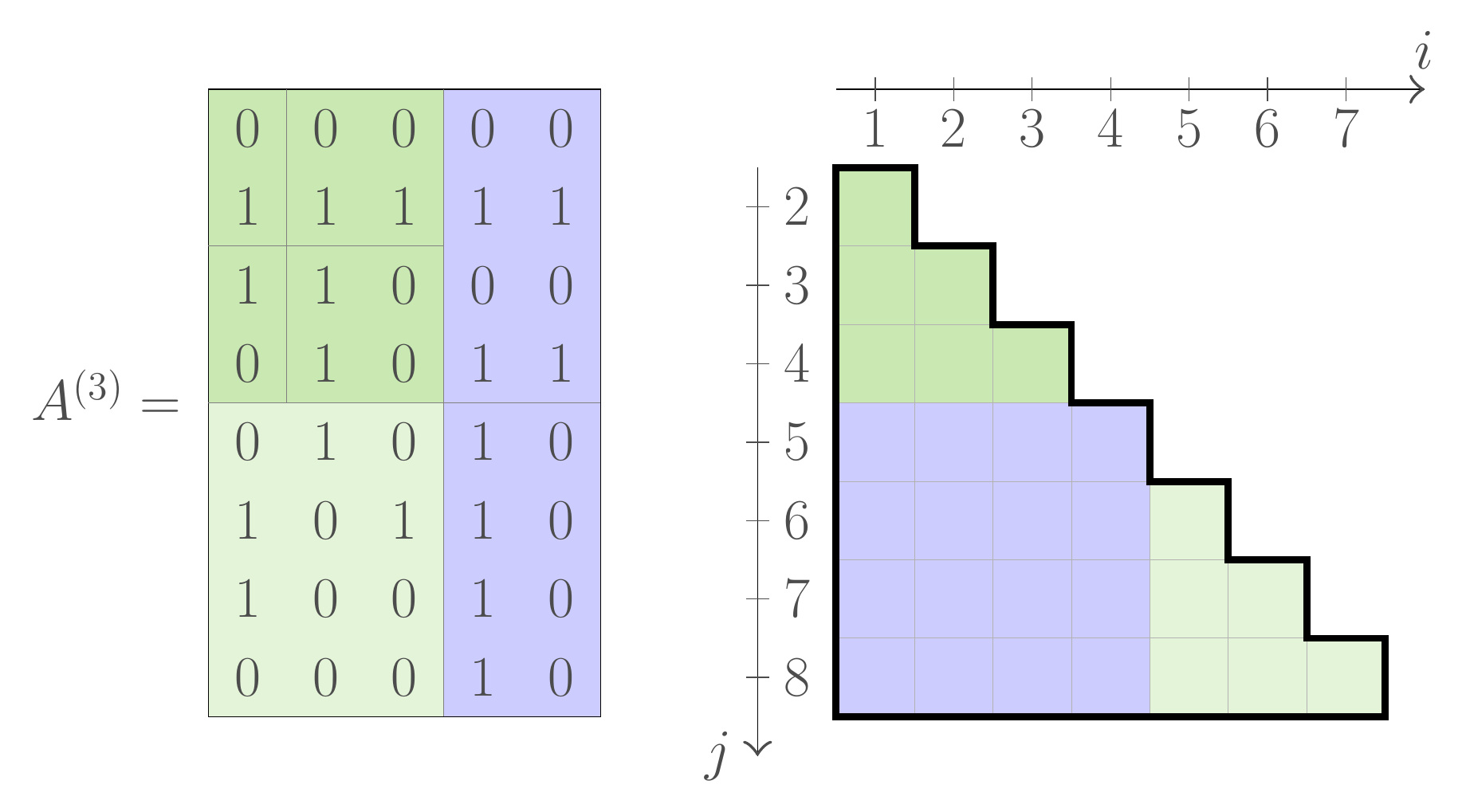}
}
\end{center}
\vspace{-.5cm}
\caption{\small One step from Construction~\ref{def:simple-construction} is illustrated. The green matrices are glued together and ensure the top and right constraints in the constraint space. The two additional blue columns take care of the remaining square.}
\label{fig:old-lower-bound-construction}
\end{figure}

\begin{lem} \label{thm:simple-construction-valid}
	Matrices obtained from Construction~\ref{def:simple-construction} are Order-Regular and satisfy $m = \Omega\big(\sqrt{2}^n\big)$ with $m$ and $n$ its number of rows and columns respectively.
	\begin{proof}
		We prove the lemma by induction on $\ell$. Clearly $A^{(1)}$ is OR (only one $(i,j)$ pair to check). We show that if $A^{(\ell-1)}$ is OR, then Order-Regularity follows for $A^{(\ell)}$. 
		
%		First, both $A^{(\ell-1)}$ and $\widetilde{A}^{(\ell-1)}$ are Order-Regular and $\widetilde{A}^{(\ell-1)}$ was defined in such a way that the gluing between $A^{(\ell-1)}$ and $\widetilde{A}^{(\ell-1)}$ is proper. Therefore, condition~\eqref{eq:OR} is ensured for every pairs $(i,j)$ that satisfy $1 \leq i < j \leq 2^{\ell-1}$ or $2^{\ell-1}+1 \leq i < j \leq 2^{\ell}$.
		First, observe that the left part of $A^{(\ell)}$ is a 2-gluing of $A^{(\ell-1)}$. Using Lemma~\ref{thm:effect-of-gluing}, we get all constraints $(i,j)$ satisfied when either $1 \leq i < j \leq 2^{\ell-1}$ or $2^{\ell-1}+1 \leq i < j \leq 2^{\ell}$. 
		The remaining constraints, i.e. those such that $1 \leq i \leq 2^{\ell-1}$ and $2^{\ell-1}+1 \leq j \leq 2^{\ell}$, are satisfied by the two last columns of $A^{(\ell)}$. Indeed, if $i$ is odd, then choosing $k$ to be the first of the two extra columns ensures condition~\eqref{eq:OR} for all $2^{\ell-1}+1 \leq j \leq 2^{\ell}$. The same goes with even $i$'s and the second of the two columns. This reasoning is illustrated by Figure~\ref{fig:old-lower-bound-construction}.
		Furthermore, the matrix $A^{(\ell)}$ satisfies $m_{\ell} = \sqrt{2}^{n_{\ell}+1}$.
	\end{proof}
\end{lem}

%In Construction~\ref{def:simple-construction}, the matrix $A^{(\ell)}$ satisfies $m_{\ell} = \sqrt{2}^{n_{\ell}+1}$. Since it is OR from Lemma~\ref{thm:simple-construction-valid}, 
Using Construction~\ref{def:simple-construction}, we have a way of building OR matrices satisfying $m=\Omega\big(\sqrt{2}^n\big)$. In the next subsections, we show how we can improve this bound.

\subsection{Building blocks}

Similarly to the above Construction~\ref{def:simple-construction}, our construction starts with a building block that we will use to trigger the recursion.
We require the following Strong Order-Regularity condition on the building block which is a restriction of the Order-Regularity condition.

\begin{defi}[Strong Order-Regularity] \label{def:SOR}
We say that $B \in \{0,1\}^{m \times n}$ is Strongly Order-Regular (SOR) whenever 
\begin{itemize}
	\item[(1)] for every pair of rows $i,j$ of $B$ with $1 \leq i < j \leq m$, there exists a column $k_1$ such that:
	\begin{align} \label{eq:SOR1}
		B_{i,k_1} \neq B_{i+1,k_1} = B_{j,k_1} = B_{j+1,k_1}
	\end{align} 
	(the original Order-Regularity condition);
	\item[(2)] for every pair of rows $i,j$ of $B$ with $1 \leq i$ and $i+1 < j \leq m$, there exists a column $k_2$ (necessarily different from $k_1$) such that:
	\begin{align} \label{eq:SOR2}
		B_{i,k_2} \neq B_{i+1,k_2} \neq B_{j,k_2} = B_{j+1,k_2}.
	\end{align} 
\end{itemize}
Again, we choose the convention that $B_{m+1,k} = B_{m,k}$ and the last two rows are required to be distinct. 
\end{defi}

In other words, at the entries $i, i+1, j, j+1$ we now ask for one column $k_1$ at which we observe either $0, 1, 1, 1$ or $1, 0, 0, 0$ and for another column $k_2$ at which we observe either $0, 1, 0, 0$ or $1, 0, 1, 1$. Clearly, this second column cannot exist when $j = i+1$, hence we do not ask for its existence in that case. We say that a matrix \emph{doubly-satisfies} a constraint $(i,j)$ if both $k_1$ and $k_2$ exist for that constraint (that is, for an SOR matrix, every constraints such that $1 \leq i$ and $i+1 < j \leq m$).
An SOR matrix with 8 columns and 33 rows is given in Figure~\ref{fig:building-block}. 

\begin{figure}[h!]
	\begin{center}
	{\large$B = B^{(1)} = $} {\footnotesize\tt
	\begin{tabular}{|cccccccc|}
	 	\hline			
	 		0&0&0&0&0&0&0&0\\
			1&1&1&1&1&1&1&1\\
			0&0&0&0&0&0&0&1\\
			0&1&1&1&1&1&1&1\\
			0&0&0&0&0&0&1&0\\
			1&0&1&1&1&1&1&1\\
			0&0&0&0&0&1&0&0\\
			1&0&0&1&1&1&1&1\\
			0&1&0&0&1&0&0&0\\
			1&0&0&1&1&1&1&0\\
			0&0&1&0&1&0&0&0\\
			1&0&0&1&1&0&1&0\\
			0&1&1&1&1&0&0&0\\
			1&0&0&1&0&0&0&1\\
			0&1&0&1&1&1&0&0\\
			1&0&0&1&0&0&0&0\\
			0&1&0&1&1&0&0&0\\
			1&0&0&1&1&0&0&1\\
			0&0&0&0&1&1&0&0\\
			0&0&0&1&1&0&1&1\\
			0&0&1&1&0&1&0&0\\
			0&0&0&1&1&0&1&0\\
			0&0&1&1&1&1&0&0\\
			0&0&0&1&1&0&0&1\\
			0&0&0&1&0&1&0&0\\
			0&0&0&1&1&1&1&1\\
			0&0&0&0&0&1&0&1\\
			0&1&0&1&0&1&1&1\\
			0&0&1&0&0&1&1&1\\
			0&1&0&0&0&1&1&1\\
			1&1&1&0&0&1&1&1\\
			0&1&1&0&0&0&1&1\\
			0&1&1&0&0&1&1&1\\
		\hline
	\end{tabular}}
	\end{center}
	\caption{\small The $33\times 8$ Strongly Order-Regular building block that we use to obtain our first improvement to the $\Omega\big(\sqrt{2}^n\big)$ lower bound. }
	\label{fig:building-block}
\end{figure}

\subsection{Blowing up}

We now provide our main construction that enables us to improve the bound. We start by describing the components of each iterate of the construction. Then we show that it indeed generates Order-Regular matrices and conclude with the resulting new lower bound.

\begin{figure}[h]
\begin{center}
\scalebox{.4}{
\pgfimage{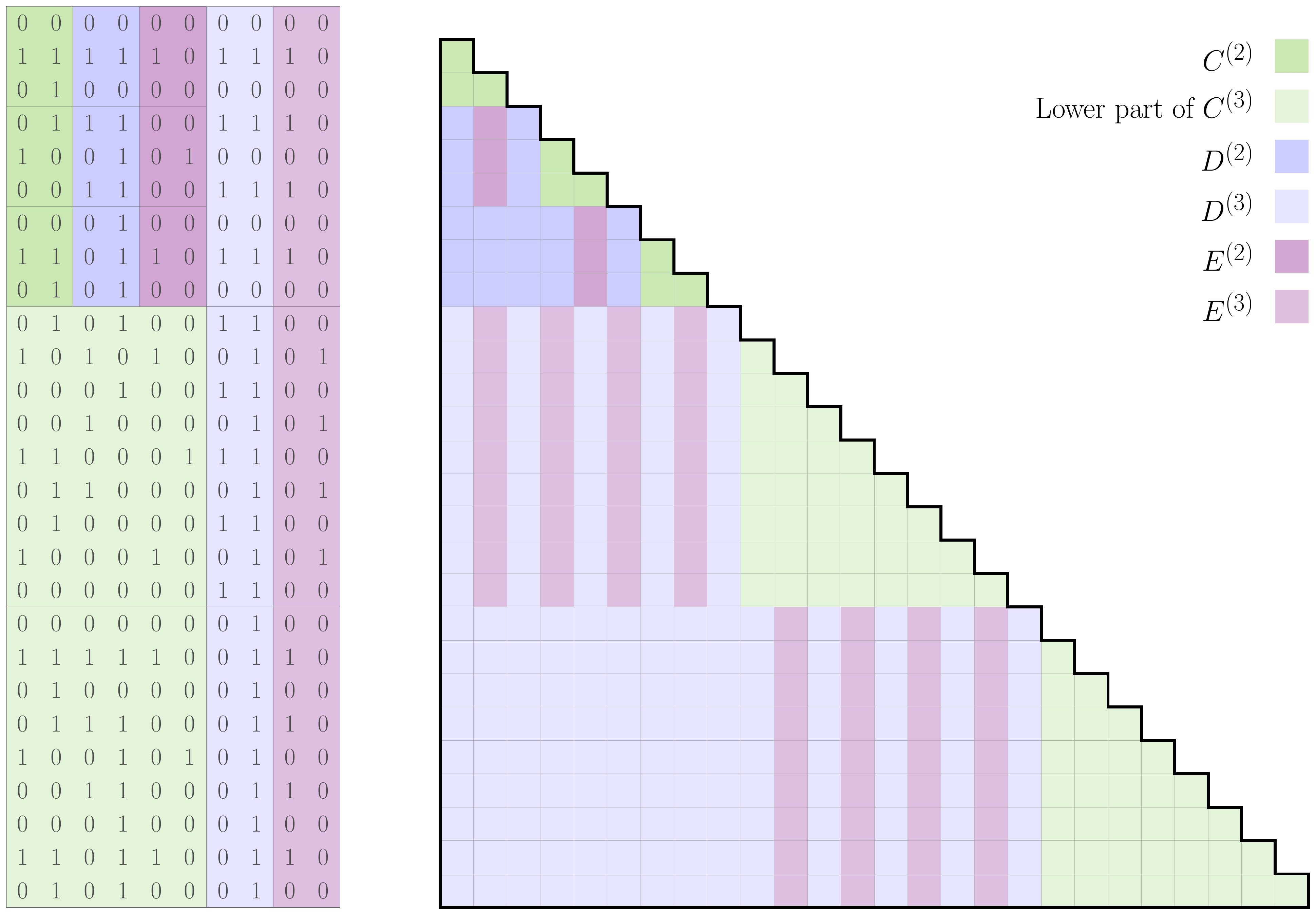}
}
\end{center}
\caption{\small An example of two blowup steps of Construction~\ref{def:main-construction} with a $3 \times 2$ SOR building block. Notice how each part of the construction contributes in filling the constraint space. The $C$ blocks (green) fill in triangles of the size of the previous iterate along the diagonal. The $D$ blocks (blue) almost fill in the rest of the space. The two last rows (violet) of each step of the construction aim to fill in the remaining holes.}
\label{fig:main-construction-example}
\end{figure}

\begin{constr} \label{def:main-construction}
	Let $B = B^{(1)} \in \{0,1\}^{M \times N}$ be an SOR matrix (the building block). We inductively build a matrix $B^{(\ell)} \in \{0,1\}^{m_{\ell} \times n_{\ell}}$ as the merging of three blocks:
	\begin{align*}
		B^{(\ell)} = \begin{bmatrix} C^{(\ell)} & D^{(\ell)} & E^{(\ell)} \end{bmatrix}.
	\end{align*}
	The blocks $C^{(\ell)}, D^{(\ell)}$ and $E^{(\ell)}$, $\ell \geq 2$, are defined as follows.
	\begin{itemize}
		\item The $C^{(\ell)}$ block is composed of $M$ copies of the previous iterate glued together:
		\begin{align*}
			C^{(\ell)}
				\triangleq \left\{ \begin{matrix} B^{(\ell-1)} \\ \widetilde{B}^{(\ell-1)} \end{matrix} \right\}_{\!M} 
				= \begin{bmatrix} B^{(\ell-1)} \\ \widetilde{B}^{(\ell-1)} \\ \vdots \\ B^{(\ell-1)} \\ \widetilde{B}^{(\ell-1)} \\ B^{(\ell-1)} \end{bmatrix}.
		\end{align*}
		\item The $D^{(\ell)}$ block expands the building block $B$ in the following way:
		\begin{align*}
			D^{(\ell)}
				\triangleq 	\begin{bmatrix} 
						d^{1,1} & d^{1,2} & \dots  & d^{1,N} \\ 
						d^{2,1} & d^{2,2} & \dots  & d^{2,N} \\ 
						\vdots  & \vdots  & \ddots & \vdots  \\
						d^{M,1} & d^{M,2} & \dots  & d^{M,N} \\
					\end{bmatrix} 
		\end{align*}
		with:
		\begin{align*}
			d^{i,k} \triangleq \left\{ \begin{matrix} B_{i,k} \\ B_{i+1,k} \end{matrix} \right\}_{\!m_{\ell-1}}
				= \begin{cases}
					\pattern{m_{\ell-1}}{0}{0} &\text{if } B_{i,k} = 0 \text{ and } B_{i+1,k} = 0 \\
					\pattern{m_{\ell-1}}{0}{1} &\text{if } B_{i,k} = 0 \text{ and } B_{i+1,k} = 1 \\
					\pattern{m_{\ell-1}}{1}{0} &\text{if } B_{i,k} = 1 \text{ and } B_{i+1,k} = 0 \\
					\pattern{m_{\ell-1}}{1}{1} &\text{if } B_{i,k} = 1 \text{ and } B_{i+1,k} = 1 
				\end{cases}
		\end{align*}
		again with the convention that $B_{M+1,k} = B_{M,k}$ for all $k$.
		\item The $E^{(\ell)}$ block is composed of two extra columns that will ensure the Order-Regularity of the whole:
		\begin{align*}
			E^{(\ell)}
				\triangleq \begin{bmatrix} 
						\left\{ \begin{matrix} \pattern{m_{\ell-1}}{0}{1} \\ \pattern{m_{\ell-1}}{0}{0} \end{matrix} \right\}_{\!M} 
						& \left\{ \begin{matrix} \pattern{m_{\ell-1}}{0}{0} \\ \pattern{m_{\ell-1}}{0}{1} \end{matrix} \right\}_{\!M} 
					\end{bmatrix}
				= \begin{bmatrix} 
					\pattern{m_{\ell-1}}{0}{1} & \pattern{m_{\ell-1}}{0}{0} \\ 
					\pattern{m_{\ell-1}}{0}{0} & \pattern{m_{\ell-1}}{0}{1} \\ 
					\vdots \hspace{.6cm} {\color{white}{.}}	& 
					\vdots \hspace{.6cm} {\color{white}{.}} \\ 
					\pattern{m_{\ell-1}}{0}{1} & \pattern{m_{\ell-1}}{0}{0} \\ 
					\pattern{m_{\ell-1}}{0}{0} & \pattern{m_{\ell-1}}{0}{1} \\ 
					\pattern{m_{\ell-1}}{0}{1} & \pattern{m_{\ell-1}}{0}{0} 
				\end{bmatrix}.
		\end{align*}
		We call $e^{i,k}$ the $i^{\text{th}}$ pattern encountered in the $k^{\text{th}}$ columns of $E^{(\ell)}$.
	\end{itemize}
	Given this construction, it follows that $m_{\ell} = M \cdot m_{\ell-1} = M^{\ell}$ and that $n_{\ell} = n_{\ell-1}+N+2 = \ell \cdot N + 2(\ell-1)$. Figure~\ref{fig:main-construction-example} helps visualizing the role of each block.
%	Notice that all three blocks are divided into $M$ \emph{slices} (that is, blocks of rows) of size $m_{\ell-1}$ each. We will say that an index $i$ belongs to a slice $s, 1 \leq s \leq M$, if $(s-1) \cdot m_{\ell-1} < i \leq s \cdot m_{\ell-1}$.
\end{constr}

\begin{defi}[Slices]
	All three blocks $C^{(\ell)}, D^{(\ell)}$ and $E^{(\ell)}$ from construction~\ref{def:main-construction} are divided into $M$ \emph{slices} (that is, blocks of consecutive rows) of size $m_{\ell-1}$ each. We say that a row index $i$ \emph{belongs to} a slice $s$, $1 \leq s \leq M$, if $(s-1) \cdot m_{\ell-1} < i \leq s \cdot m_{\ell-1}$. We also say that $i$ corresponds to an \emph{odd} (or \emph{even}) index of $s$ if 
	its relative index within $s$ is odd (or even).
%	it corresponds to the $x^{\text{th}}$ row with $x$ an odd (resp. even) number.
\end{defi}

We now prove the central lemma of this section.

\begin{lem} \label{thm:main-construction-valid}
	Matrices obtained from Construction~\ref{def:main-construction} using a Strongly Order-Regular matrix $B$ with an odd number of rows as building block are Order-Regular.
	\begin{proof}
		Clearly, $B^{(1)} = B$ is OR because it is also Strongly OR. Assuming that $B^{(\ell-1)}$ is OR, let us show that $B^{(\ell)}$ is also OR. Therefore, we show that each block of the construction is designed to satisfy complementary subsets of the constraints space. Figure~\ref{fig:main-construction-example} graphically illustrates on a particular case how each block contributes in filling in the constraint space.
		
		\vspace{.1cm}
		\textbf{Claim 1.} \emph{The $C^{(\ell)}$ block satisfies every constraint $(i,j)$ where $i$ and $j$ belong to the same slice $s$ of $B^{(\ell)}$.}
		\vspace{.1cm}
%		\boldmath\textbf{Claim 1. The $C^{(\ell)}$ block satisfies every constraint $(i,j)$ where $i$ and $j$ belong to the same slice $s$ of $B^{(\ell)}$.}\unboldmath
%		{\color{black!50}{, that is, where $(s-1) \cdot m_{\ell-1}+1 \leq i < j \leq s \cdot m_{\ell-1}$ for some integer $1 \leq s \leq M$.}}}\unboldmath
		
		Claim 1 follows directly from Lemma~\ref{thm:effect-of-gluing} since $C^{(\ell)}$ is simply an $M$-gluing of $B^{(\ell-1)}$ which has $m_{\ell-1}$ rows by definition.
				
		\vspace{.1cm}
		\textbf{Claim 2.} \emph{The $D^{(\ell)}$ block satisfies every constraint $(i,j)$ where $i$ and $j$ belong to two different and non-adjacent slices $s_i$ and $s_j$.}
		\vspace{.1cm}
%		{\color{black!50}{, that is, where $(s_i-1) \cdot m_{\ell-1} < i \leq s_i \cdot m_{\ell-1}$ and $(s_j-1) \cdot m_{\ell-1} < j \leq s_j \cdot m_{\ell-1}$ for some integers $1 \leq s_i < s_j \leq M$ such that $s_j > s_i+1$.}}}\unboldmath
		
		Let $(i,j)$ be such a constraint for some integers $s_i$ and $s_j$. From the Strong Order-Regularity of $B$ and the fact that $s_i$ and $s_j$ are non adjacent (that is, $s_i + 1 < s_j$), we know that $B$ doubly-satisfies the constraint $(s_i, s_j)$. Therefore, from the definition of $D^{(\ell)}$, we know that there exist two columns $k_1$ and $k_2$ of $D^{(\ell)}$ such that the patterns that appear in the slices $s_i$ and $s_j$ for these columns are of the form:
		\begin{align*}
			d^{s_i,k_1} &= \pattern{}{\overline{\alpha}}{\alpha} & d^{s_i,k_2} &= \pattern{}{\beta}{\overline{\beta}}\\
			d^{s_j,k_1} &= \pattern{}{          \alpha }{\alpha} & d^{s_j,k_2} &= \pattern{}{\beta}{\beta}
		\end{align*}
%		$d^{s_i,k_1} = \pattern{}{\alpha}{\beta}$, $d^{s_j,k_1} = \pattern{}{\beta}{\beta}$, $d^{s_i,k_2} = \pattern{}{\gamma}{\delta}$ and $d^{s_j,k_2} = \pattern{}{\gamma}{\gamma}$ 
		for some $\alpha, \beta \in \{0,1\}$ where $\overline{\alpha} = 1-\alpha$ and $\overline{\beta} = 1-\beta$. Let $I(i,j) \triangleq \begin{bmatrix} i & i+1 & j & j+1 \end{bmatrix}$ be the vector of row indices needed when checking the OR condition~\eqref{eq:OR} for the constraint $(i,j)$. Then, using Matlab-like notations, two cases are possible: 
		\begin{itemize}
			\item either $D^{(\ell)}_{I(i,j),k_1} = \begin{bmatrix} \overline{\alpha} & \alpha & \alpha & \alpha \end{bmatrix}$ and $D^{(\ell)}_{I(i,j),k_2} = \begin{bmatrix} \beta & \overline{\beta} & \beta & \beta \end{bmatrix}$;
			\item or $D^{(\ell)}_{I(i,j),k_1} = \begin{bmatrix} \alpha & \overline{\alpha} & \alpha & \alpha \end{bmatrix}$ and $D^{(\ell)}_{I(i,j),k_2} = \begin{bmatrix}  \overline{\beta} & \beta & \beta & \beta \end{bmatrix}$.
		\end{itemize}
		In one or the other case there will always be a column $k$, either $k_1$ or $k_2$, such that condition~\eqref{eq:OR} is verified. This will be true even if $i+1$ or $j+1$ belong to the next slice (respectively $s_i+1$ or $s_j+1$) thanks to the assumption that the building block $B$, and therefore also every iterate of Construction~\ref{def:main-construction}, have an odd number of rows. Indeed, because of this parity, any pattern $d^{i,k}$, of the form $\pattern{}{\alpha}{\beta}$, ends with an $\alpha$ and the next pattern below it starts over with a $\beta$, thereby continuing the alternation of $\alpha$ and $\beta$ for one more row.
			
		\vspace{.1cm}			
		\textbf{Claim 3.} \emph{The $D^{(\ell)}$ block also satisfies every constraint $(i,j)$ where $i$ and $j$ belong to two adjacent slices $s$ and $s+1$ and $i$ corresponds to an \emph{odd} index of $s$.}
		\vspace{.1cm}
%		{\color{black!50}{, that is, where $(s-1) \cdot m_{\ell-1} < i \leq s \cdot m_{\ell-1} < j \leq (s+1) \cdot m_{\ell-1}$ for some integer $1 \leq s < M$ and $i - (s-1) \cdot m_{\ell-1}$ is an \emph{odd} number.}}}\unboldmath
		
		In the case of adjacent slices, condition~\eqref{eq:SOR2} is no longer ensured for $B$. However, the original Order-Regularity still holds and there exists a column $k$ of $D^{(\ell)}$ such that $d^{s,k} = \pattern{}{\overline{\alpha}}{\alpha}$ and $d^{s+1,k} = \pattern{}{\alpha}{\alpha}$ for some $\alpha \in \{0,1\}$. Since $i$ corresponds to an odd index of $s$, we must have $D^{(\ell)}_{i,k} = \overline{\alpha}$ and therefore we have $D^{(\ell)}_{I(i,j),k} = \begin{bmatrix} \overline{\alpha} & \alpha & \alpha & \alpha \end{bmatrix}$ which confirms Claim~3.
		
		\vspace{.1cm}
		\textbf{Claim 4.} \emph{The $E^{(\ell)}$ block satisfies every constraint $(i,j)$ where $i$ and $j$ belong to two adjacent slices $s$ and $s+1$ and $i$ corresponds to an \emph{even} index of $s$.}
		\vspace{.1cm}
%		{\color{black!50}{, that is, where $(s-1) \cdot m_{\ell-1} < i \leq s \cdot m_{\ell-1} < j \leq (s+1) \cdot m_{\ell-1}$ for some integer $1 \leq s < M$ and $i - (s-1) \cdot m_{\ell-1}$ is an \emph{even} number.}}}\unboldmath
		
%		Let $e^{s,k}$ refer to the $k^{\text{th}}$ column of the slice $s$ in the block $E^{(\ell)}$. 
		From the definition of $E^{(\ell)}$, we know that there is always one of the two columns, say $k$, such that $e^{s,k} = \pattern{}{0}{1}$ and $e^{s+1,k} = \pattern{}{0}{0}$. Since $i$ corresponds to an even index of $s$, it means that $E^{(\ell)}_{i,k} = 1$ and therefore we have $E^{(\ell)}_{I(i,j),k} = \begin{bmatrix} 1 & 0 & 0 & 0 \end{bmatrix}$ which confirms Claim~4.
		
		\vspace{.1cm}
		\textbf{Summary.} Given any constraint $(i,j)$:
		\begin{itemize}
			\item if $i$ and $j$ belong to the same slice, then the Order-Regularity condition holds for the constraint from Claim~1;
			\item if they belong to different slices that are non-adjacent to each other, then the condition holds from Claim~2;
			\item if they belong to adjacent slices, then the condition holds from Claims~3 and~4 together;
		\end{itemize}
		Therefore, all constraints are satisfied by $B^{(\ell)}$.
	\end{proof}
\end{lem}

\begin{prop}[A first improvement on the lower bound] \label{thm:first-bound}
%	It is possible to build Order-Regular matrices with $n$ columns and $\Omega(\sqrt[10]{33}^n) = \Omega(1.4186^n)$ rows.
	For all $n$ there exists an $n$-column Order-Regular matrix  with at least $m = \big(\!\sqrt[10]{33}\,\big)^{n-7} = \Omega(1.4186^n)$ rows.
	
	\begin{proof}
		We use Construction~\ref{def:main-construction} with the $33 \times 8$ building block from Figure~\ref{fig:building-block}. After $\ell$ steps of the construction, we get a matrix $B^{(\ell)}$ with $m = 33^{\ell}$ rows and $n = 10\ell-2$ columns and therefore $m = 33^{(n+2)/10} = \Omega(\sqrt[10]{33}^n)$ when $n = 8, 18, 28, \ldots$ From Lemma~\ref{thm:main-construction-valid}, this matrix is Order-Regular. For a value of $n$ such that $10\ell-2 < n < 10(\ell+1)-2$ for some integer $\ell$, the same construction as for $n = 10\ell-2$ applies (simply add up to $9$ dummy columns to the construction to match the required number of columns). Here the worst case is when $n = 10(\ell+1)-3$ and this is why we subtracted $9$ in the exponent of the bound such that it holds for any value of $n$, with no incidence on the rate of growth. 
%		Asymptotically when $\ell$ goes to infinity, we generate Order-Regular matrices with $n$ columns and $\Omega(\sqrt[10]{33}^n)$ rows.
	\end{proof}
\end{prop}

\subsection{One step further: modified building block constraints}

\begin{defi}[Partially-Strong Order-Regularity] \label{def:PSOR}
We say that $B \in \{0,1\}^{m \times n}$ is Partially-Strongly Order-Regular (PSOR) whenever 
\begin{itemize}
	\item[(1)] for every pair of rows $i,j$ of $B$ with $1 \leq i < j \leq m$, there exists a column $k_1$ such that:
	\begin{align*}
		B_{i,k_1} \neq B_{i+1,k_1} = B_{j,k_1} = B_{j+1,k_1}
	\end{align*} 
	(the original Order-Regularity condition);
	\item[(2)] for every pair of rows $i,j$ of $B$ with $1 < i < j < m$ and for which $j-i$ is even, there exists a column $k_2$ (necessarily different from $k_1$) such that:
	\begin{align*}
		B_{i,k_2} \neq B_{i+1,k_2} \neq B_{j,k_2} = B_{j+1,k_2}.
	\end{align*} 
\end{itemize}
Once again, we choose the convention that $B_{m+1,k} = B_{m,k}$ and the last two rows are required to be distinct. 
\end{defi}

\begin{figure}[h!]
\begin{center}
\scalebox{.224}{
\pgfimage{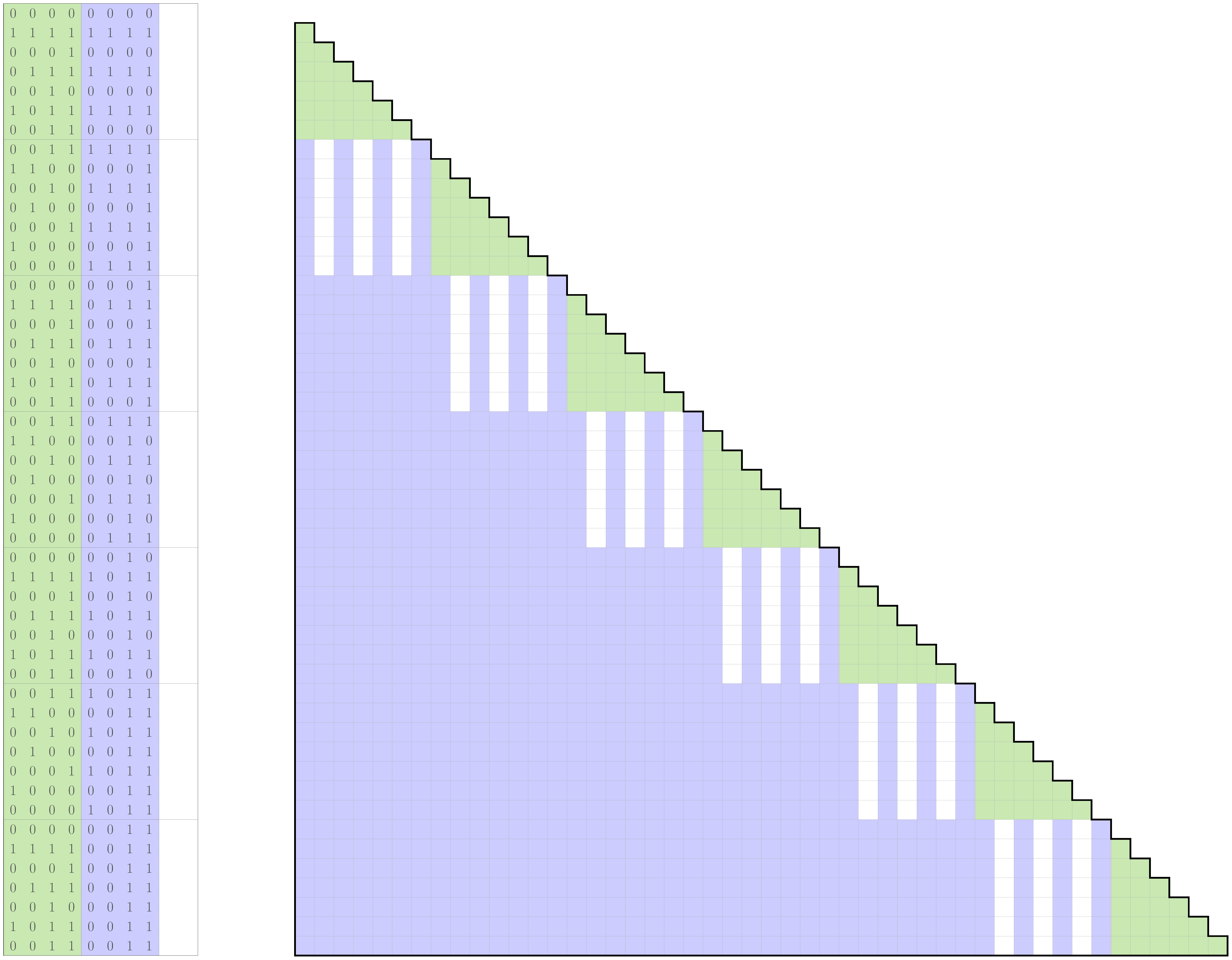}
}
\end{center}

\begin{center}
\scalebox{.224}{
\pgfimage{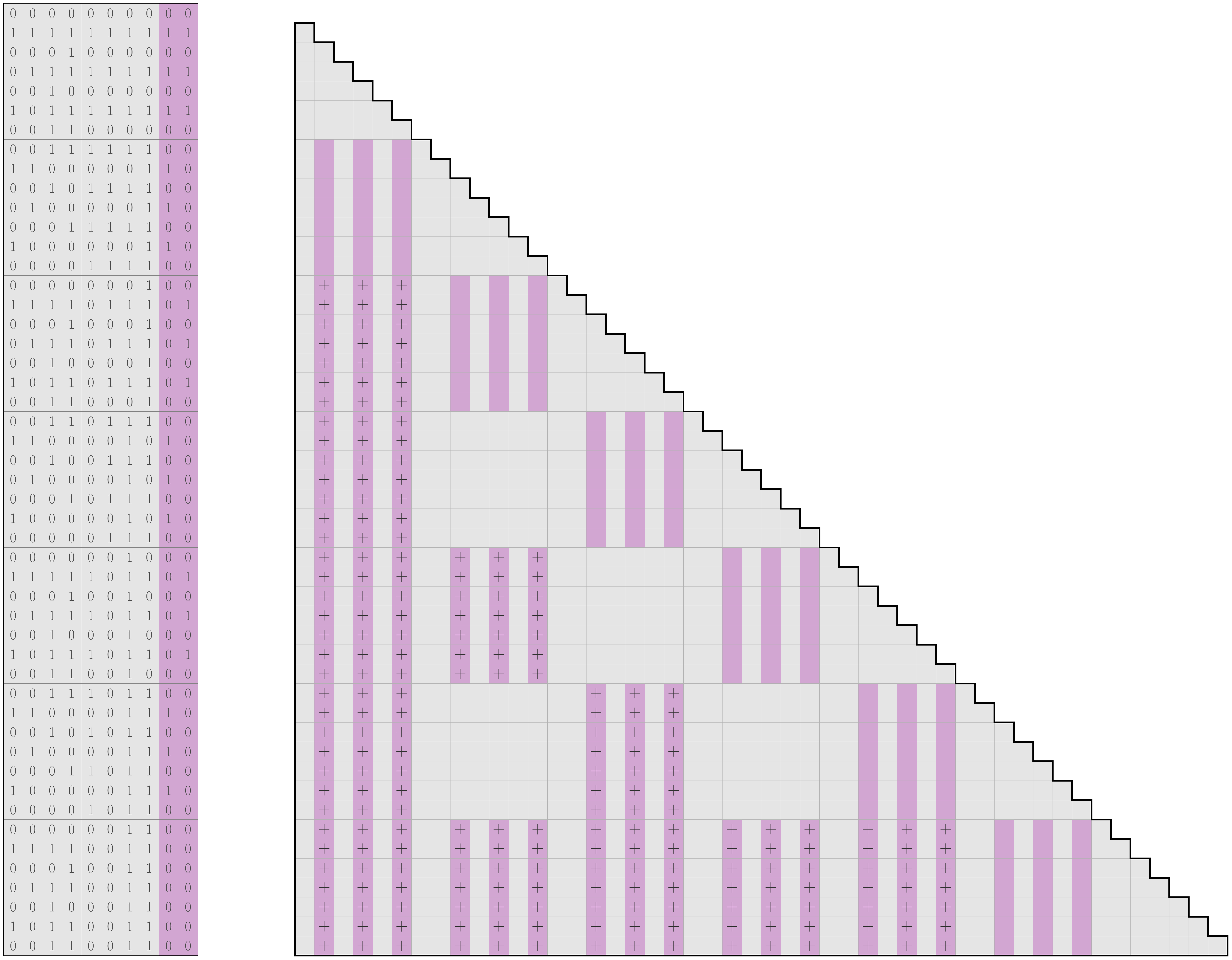}
}
\end{center}
\caption{\small If we only used the $C^{(\ell)}$ and $D^{(\ell)}$ blocks (respectively in green and blue) in Construction~\ref{def:main-construction} with an SOR building block, there would remain a few holes in the constraint space (top). The $E^{(\ell)}$ block is designed to fill these holes. But a slightly improved version of the $E^{(\ell)}$ block (violet) actually fills much more that just the required holes (bottom). This fact can be exploited to soften the constraints on the building block and further improve the lower bound to obtain our final bound.}
\label{fig:effect-of-filling-holes-example}
\end{figure}

The difference with the Strong Order-Regularity lies in the second condition. We now no longer require the existence of the column $k_2$ when $i = 1$, when $j = m$ or when $j-i$ is an odd number, hence the constraints that are doubly-satisfied by a PSOR matrix are those such that $1 < i < j < m$ and $j-i$ is even. As illustrated by Figure~\ref{fig:effect-of-filling-holes-example}, we are allowed to do this relaxation because the $E^{(\ell)}$ block from Construction~\ref{def:main-construction} actually satisfies more constraints than the sole ones it was designed to satisfy initially (as referred to in Claim~4 of the proof of Lemma~\ref{thm:main-construction-valid}) making it possible to reduce the set of constraints that the $D^{(\ell)}$ block has to satisfy and hence to soften the SOR condition. The softened condition allows us to find a larger building block which in turn results in an improved lower bound.
%But the actual improvement to the bound from Proposition~\ref{thm:first-bound} comes from the fact that we can now use a strictly larger building block, given in Figure~\ref{fig:softened-building-block}.

\begin{figure}[h!]
	\begin{center}
	{\large$\myhat{B} = \myhat{B}^{(1)} = $} {\footnotesize\tt
	\begin{tabular}{|cccccccc|}
	 	\hline			
			0&0&0&0&0&0&0&0\\
			1&1&1&1&1&1&1&1\\
			0&0&0&0&0&0&0&1\\
			0&1&1&1&1&1&1&1\\
			0&0&0&0&0&0&1&0\\
			1&0&1&1&1&1&1&1\\
			1&0&0&0&0&1&0&0\\
			0&1&0&1&1&1&1&1\\
			1&1&0&0&0&0&0&1\\
			1&1&0&1&1&1&1&0\\
			1&0&0&0&1&0&0&1\\
			0&1&1&0&1&1&1&0\\
			1&0&0&0&1&0&0&0\\
			0&1&1&1&1&1&0&0\\
			0&0&0&0&1&0&0&0\\
			0&1&1&1&1&0&0&1\\
			1&1&0&1&0&0&0&0\\
			0&0&1&1&1&0&1&0\\
			0&1&1&0&0&0&0&0\\
			1&1&1&1&1&0&1&0\\
			0&1&1&0&1&0&0&0\\
			1&1&0&0&1&0&1&0\\
			1&1&1&0&1&0&0&1\\
			1&0&1&0&1&1&1&0\\
			1&1&1&0&1&0&0&0\\
			1&1&1&0&0&1&1&0\\
			1&1&1&0&1&0&1&1\\
			1&0&1&0&0&0&1&0\\
			0&1&1&0&1&0&1&0\\
			0&1&1&1&0&0&1&0\\
			0&0&1&0&0&0&1&1\\
			0&1&0&0&0&0&1&1\\
			1&1&1&0&0&0&1&1\\
			0&1&1&0&0&0&1&0\\
			0&1&1&0&0&0&1&1\\
		\hline
	\end{tabular}}
	\end{center}
	\caption{\small The $35\times 8$ Partially-Strongly Order-Regular building block that we use to obtain our final lower bound. }
	\label{fig:softened-building-block}
\end{figure}

Before we show why using PSOR building blocks results in OR matrices, we need to slightly adapt Construction~\ref{def:main-construction}, and more precisely the definition of the $E^{(\ell)}$ block.

\addtocounter{myconstr}{+1}
\begin{myconstr}
	Let $\myhat{B} = \myhat{B}^{(1)} \in \{0,1\}^{M \times N}$ be a PSOR building block matrix. In the same spirit as Construction~\ref{def:main-construction}, we inductively build a matrix $\myhat{B}^{(\ell)} \in \{0,1\}^{m_{\ell} \times n_{\ell}}$ as the merging of three blocks:
	\begin{align*}
		\myhat{B}^{(\ell)} = \begin{bmatrix} \myhat{C}^{(\ell)} & \myhat{D}^{(\ell)} & \myhat{E}^{(\ell)} \end{bmatrix},
	\end{align*}
	where the definitions of $\myhat{C}^{(\ell)}$ and $\myhat{D}^{(\ell)}$ are the same as those of $C^{(\ell)}$ and $D^{(\ell)}$ in Construction~\ref{def:main-construction} with $\myhat{B}^{(\ell)}$ and $\myhat{B}$ taking the role of $B^{(\ell)}$ and $B$ respectively and where 
	\begin{align*}
		\myhat{E}^{(\ell)} = \begin{bmatrix} 
			\myhat{e}^{1,1} & \myhat{e}^{1,2} \\ 
			\myhat{e}^{2,1} & \myhat{e}^{2,2} \\ 
			\vdots 					   & \vdots \\ 
			\myhat{e}^{M,1} & \myhat{e}^{M,2} \\ 
		\end{bmatrix}
	\end{align*}
	is a slight modification of $E^{(\ell)}$ such that:
	\begin{align*}
		\myhat{e}^{i,k} = \begin{cases}
			\pattern{}{0}{1} & \text{if } i=1 \text{ and } k=2,\\
			\pattern{}{0}{0} & \text{if } i=M \text{ and } k=1,\\
			e^{i,k} & \text{otherwise.}
		\end{cases}
	\end{align*}

%	But first, observe in the first column of the first slice of $E^{(\ell)}$ that starting with a vector of zeros $\pattern{}{0}{0}$ is pointless as it does not help to satisfy any constraint. Therefore, we replace this vector with a vector of switches $\pattern{}{0}{1}$ of the same dimension. The same observation holds for the last slice of the second column: we replace the vector of switches $\pattern{}{0}{1}$ by a vector of zeros $\pattern{}{0}{0}$. We call $\widetilde{E}^{(\ell)}$ the resulting block that replaces $E^{(\ell)}$. As in the Claim~4 of the proof of Lemma~\ref{thm:main-construction-valid}, let us denote by $\tilde{e}^{s,k}$ the $k^{\text{th}}$ column of the slice $s$ of $\widetilde{E}^{(\ell)}$.
\end{myconstr}

The only changes compared to Construction~\ref{def:main-construction} is that the second column of $\myhat{E}^{(\ell)}$ now starts with $\pattern{}{0}{1}$ instead of $\pattern{}{0}{0}$ and its first column now ends with $\pattern{}{0}{0}$ instead of $\pattern{}{0}{1}$. Clearly, the modification can only help to satisfy more constraints.

\begin{lem} \label{thm:softened-construction-valid}
	Matrices obtained from Construction~\ref{def:main-construction}$\;\!^*$ using Partially-Strongly Order-Regular building blocks with an odd number of rows are Order-Regular.
	
	\begin{proof}
	
		The proof is inductive, in the same flavor as the proof of Lemma~\ref{thm:main-construction-valid}. Knowing that $\myhat{B}^{(1)}$ is OR and assuming that $\myhat{B}^{(\ell-1)}$ is OR too, we show that $\myhat{B}^{(\ell)}$ must also be OR.
	
		\vspace{.1cm}
		\textbf{Claim 1.} \emph{The $\myhat{C}^{(\ell)}$ block satisfies every constraint $(i,j)$ where $i$ and $j$ belong to the same slice $s$ of $\myhat{B}^{(\ell)}$.}
		\vspace{.1cm}
		
		The argument is the same as for Claim~1 in the proof of Lemma~\ref{thm:main-construction-valid}.
				
		\vspace{.1cm}
		\textbf{Claim 2.} \emph{The $\myhat{D}^{(\ell)}$ block satisfies every constraint $(i,j)$ where $i$ and $j$ belong to two different slices $s_i$ and $s_j$ such that $s_i \neq 1$, $s_j \neq M$ and $s_j - s_i$ is even.}
		\vspace{.1cm}
		
		From the Partially-Strong Order-Regularity of the building block $\myhat{B}$ and the conditions on $s_i$ and $s_j$, we know that the constraints $(s_i, s_j)$ is doubly-satisfied by $\myhat{B}$. Therefore, the same reasoning as the one of Claim~2 in the proof of Lemma~\ref{thm:main-construction-valid} applies.
					
		\vspace{.1cm}	
		\textbf{Claim 3.} \emph{The $\myhat{D}^{(\ell)}$ block also satisfies every constraint $(i,j)$ where $i$ and $j$ belong to two different slices $s_i$ and $s_j$ such that $s_i = 1$, $s_j = M$ or $s_j - s_i$ is odd and such that $i$ corresponds to an \emph{odd} index of $s_i$.}
		\vspace{.1cm}
		
		Here, the constraint $(s_i, s_j)$ is not doubly-satisfied by $\myhat{B}$ but $i$ corresponds to an \emph{odd} index of $s_i$. Again, the same argument as for Claim~3 in the proof of Lemma~\ref{thm:main-construction-valid} applies here.
		
		\vspace{.1cm}
		\textbf{Claim 4.} \emph{The $\myhat{B}^{(\ell)}$ block satisfies every constraint $(i,j)$ where $i$ and $j$ belong to two different slices $s_i$ and $s_j$ such that $s_i = 1$, $s_j = M$ or $s_j - s_i$ is odd and such that $i$ corresponds to an \emph{even} index of $s_i$.}
		\vspace{.1cm}
	
		We evaluate the three possible cases when $i$ corresponds to an even index of $s_i$.
		\begin{enumerate}[(1)]
			\item If $s_i = 1$, we have $\myhat{E}^{(\ell)}_{[\begin{smallmatrix} i & i+1 \end{smallmatrix}],k} = \begin{bmatrix} 1 & 0 \end{bmatrix}$ for both columns $k = 1$ and $2$ (since $i$ corresponds to an even index of $s_i$), and we have $\myhat{E}^{(\ell)}_{[\begin{smallmatrix} j & j+1 \end{smallmatrix}],k} = \begin{bmatrix} 0 & 0 \end{bmatrix}$ for either $k=1$ or $k=2$.
			\item When $s_j = M$, we have $\myhat{E}^{(\ell)}_{[\begin{smallmatrix} j & j+1 \end{smallmatrix}],k} = \begin{bmatrix} 0 & 0 \end{bmatrix}$ for both $k = 1$ and $2$, and we have $\myhat{E}^{(\ell)}_{[\begin{smallmatrix} i & i+1 \end{smallmatrix}],k} = \begin{bmatrix} 1 & 0 \end{bmatrix}$ for either $k=1$ or $k=2$.
			\item If $s_i \neq 1, s_j \neq M$ and $s_j-s_i$ is an odd number, then $\myhat{e}^{s_i,k_1} = \pattern{}{0}{0}$ and $\myhat{e}^{s_i,k_2} = \pattern{}{0}{1}$ or vice versa. Furthermore, $\myhat{e}^{s_i,k}$ and $\myhat{e}^{s_j,k}$ are different patterns for both $k=1$ and $2$ (either $\pattern{}{0}{0}$ and $\pattern{}{0}{1}$ or vice versa). Therefore, there will always be one of the two columns, say $k'$, such that $\myhat{e}^{s_i,k'} = \pattern{}{0}{1}$ and $\myhat{e}^{s_j,k'} = \pattern{}{0}{0}$ and hence such that $\myhat{E}^{(\ell)}_{I(i,j),k'} = \begin{bmatrix} 1 & 0 & 0 & 0 \end{bmatrix}$.
		\end{enumerate}
		
		\vspace{.1cm}
		\textbf{Summary.} A constraint $(i,j)$ such that $i$ and $j$ belong to the slices $s_i$ and $s_j$ is satisfied by:
		\begin{itemize}
			\item the $\myhat{C}^{(\ell)}$ block if $s_i = s_j$;
			\item the $\myhat{D}^{(\ell)}$ block if $s_i \neq s_j$ and the constraint $(s_i, s_j)$ is doubly-satisfied by $B$;
			\item either the $\myhat{D}^{(\ell)}$ block or the $\myhat{E}^{(\ell)}$ block if $s_i \neq s_j$ and the constraint $(s_i, s_j)$ is not doubly-satisfied by $B$ (which is the case when $s_i = 1$, $s_j = M$ or $s_j - s_i$ is odd). \qedhere
		\end{itemize}

	\end{proof}
\end{lem}

\begin{thm} \label{thm:main-bound}
%	It is possible to build Order-Regular matrices of size $\Omega(\sqrt[10]{35}^n) = \Omega(1.4269^n)$.
	Given a number of columns $n$, there exists an Order-Regular matrix with at least $m = \big(\!\sqrt[10]{35}\,\big)^{n-7} = \Omega(1.4269^n)$ rows.
		
	\begin{proof}
		The proof is analog to the one of Proposition~\ref{thm:first-bound} with the PSOR building block from Figure~\ref{fig:softened-building-block} and using Lemma~\ref{thm:softened-construction-valid} to guarantee that the construction indeed provides OR matrices.
	\end{proof}
\end{thm}

\section{Techniques for building large matrices} \label{sec:large-matrices}

Our results heavily rely on our ability to build large (PS)OR matrices efficiently. First, to disprove Conjecture~\ref{thm:hansen-zwick-conjecture}, we performed an exhaustive search on the massive set of OR matrices with $n=7$ and found no matrix with $34 = F_{n+2}$ rows. Then, to obtain our lower bounds in Proposition~\ref{thm:first-bound} and Theorem~\ref{thm:main-bound}, we searched for large enough matrices in the even huger set of (P)SOR matrices with $n=8$.

Let us illustrate the size of the search space. First regarding the exhaustive search, $3\times10^{11}$ is a conservative lower bound on the total number of OR matrices with $n=7$, excluding symmetrical cases\footnote{\footnotesize We extrapolate the exact number to be around $3\times10^{16}$ using a doubly exponential regression from the number of branches for $n=1$ to $6$.}. Therefore, we cannot afford to examine each of these matrices individually and performing an exhaustive search requires to come up with some additional tricks. Furthermore, the size of the search space grows doubly exponentially with $n$ hence stepping from $n$ to $n+1$ columns makes a big difference. 
%To emphasize this fact, our final code took 1 month to run for $n=7$ for the exhaustive search while it finished in 10 seconds for $n=6$ (using 10 Intel\textregistered\ Xeon\textregistered\ X5670 cores). 
Including all the tricks and optimization described below, we were able to reduce the execution time to 1 month for $n=7$ (using 10 Intel\textregistered\ Xeon\textregistered\ X5670 cores). As a comparison, the final code took less than $10$ seconds for $n=6$. This time increase when incrementing $n$ by one suggests that the exhaustive search for $n=8$ is very challenging.
Regarding the search for building blocks, the total number of (P)SOR matrices with $n=8$ is significantly larger than that of OR matrices with $n=7$. However in that case, we only need to find one matrix that is as large as possible, which we achieve through the design of an efficient search strategy.

In the rest of this section, we present the techniques that we used to search the space of OR matrices without having to scan every candidate matrix and provide a pseudo-code of our algorithm. We also present the specific ideas that we used to perform an exhaustive search on the space and describe our search strategy to look for large matrices when an exhaustive search is neither within reach, nor necessary.

\subsection{General principles} \label{sec:large-matrices-general}

The steps below focus on OR matrices but an equivalent procedure applies for (P)SOR matrices as well.

{\bf Symmetry.} OR matrices stay OR when a permutation or a negation is applied to some of their columns. Therefore we always assume that the columns follow each other in a lexicographical order and that the first row is composed of all 0 entries. We can also assume that the second row is composed of all 1 entries since starting a column with, e.g., 00, can only satisfy less constraints than the same (negated) column that would start with 01 instead. This way we remove redundancy in the search space.

{\bf Branching.} If the first block of $d$ rows of a matrix is infeasible itself there is no need to check the rest of the matrix. On the other hand, if the first $d$ rows of several matrices are the same, it is unnecessary to recheck this part every time. We exploit these observations by using a depth first search on the matrices. If we have an initial block of $d$ rows that is feasible, we try every extension to $d+1$ rows and only continue with those that do not violate the OR condition. We are thus exploring a huge search tree whose root is by default the empty matrix and for which any node at \emph{depth} $d$, that is at distance $d$ from the root, corresponds to an OR matrix of size $d \times n$. 
%At any time of the search, we only store information about the current branch (that is, the current node $v$ and every node that stand between $v$ and the root) to allow backtracking. Therefore, we only need a limited amount of memory.

\begin{rem}[Order-Regularity$^*$] \label{rem:variation-OR}
	In this section, we use a variation of the OR condition which we refer to as OR$^*$: we require that there exists a column $k$ such that identity~\eqref{eq:OR} is verified for all $1 \leq i < j < m$ but not for $j = m$ and we allow the last two rows to be equal. Both conditions are equivalent. Indeed, from an OR$^*$ matrix, remove the last row and it becomes OR. On the other hand, take an OR matrix and copy its last row to obtain an OR$^*$ matrix. Therefore, there exists an OR matrix with $m$ rows iff there also exists an OR$^*$ matrix with $m+1$ rows. Similarly, we refer to the same variation of the (P)SOR condition by (P)SOR$^*$.
\end{rem}

{\bf Filtering.} During the branch search, assume we are investigating a branch with the first $d$ rows fixed. In any extension, any pair of rows that we encounter later has to be compatible with the same first $d$ rows. We can see these pairs as the rows labeled $j$ and $j+1$ in the order-regularity condition, to be compared with the pairs labeled $i$ and $i+1$ with $i<d$. 

\begin{defi}[Compatible pairs]
	Let $A$ be some $d \times n$ Order-Regular$^*$ matrix. We define $P_A$, the \emph{set of compatible pairs} of $A$, as:
	\begin{align} \label{eq:compatible-pairs}
		P_A = \Big\{ &(r, q) : r,q \in \{0,1\}^n \text{ and } \forall\,i,\, 1 \leq i < d, \exists\,k,\, 1 \leq k \leq n \text{ such that } A_{i,k} \neq A_{i+1,k} = r_k = q_k \Big\}.
	\end{align}
	We also define $R_A$ and $Q_A$, the projections of $P_A$ on the set of rows that respectively appear as the first and second entry of a pair:
	\begin{align}
		R_A &= \Big\{ r \in \{0,1\}^n : \exists\,q\, \in \{0,1\}^n \text{ for which } (r, q) \in P_A \Big\},  \label{eq:r-projection}\\
		Q_A(r) &= \Big\{ q \in \{0,1\}^n : (r, q) \in P_A \Big\}. \label{eq:q-projection}
	\end{align}
	
	Figure~\ref{fig:PRQ} illustrates how $P_A$, $R_A$ and $Q_A$ relate to each other on an example matrix.

\begin{figure}[h!]
\begin{center}
\begin{tikzpicture}
	% A
	
	\node[draw=none] at (-6.5,.7) {{\Large $A =$}};
	\node[draw=none] at (-5.1,1.1) {0};
	\node[draw=none] at (-5.4,1.1) {0};
	\node[draw=none] at (-5.7,1.1) {0};
	\node[draw=none] at (-5.1, .7) {1};
	\node[draw=none] at (-5.4, .7) {1};
	\node[draw=none] at (-5.7, .7) {1};
	\node[draw=none] at (-5.1, .3) {1};
	\node[draw=none] at (-5.4, .3) {0};
	\node[draw=none] at (-5.7, .3) {1};
	\draw[black] (-5.9,0) rectangle (-4.9,1.4);
	
	\node[draw=none] at (-3.75,.7) {{\Large $\Rightarrow$}};

	% P_A
	
	\node[draw=none] at (-2.1, .7) {{\Large $P_A =$}};
				
	\draw[line width=.8mm,decorate,decoration={brace,amplitude=8pt}] (-1,-.4) -- (-1,1.8);
	\draw[line width=.8mm,decorate,decoration={brace,amplitude=8pt,mirror}] (5.8,-.4) -- (5.8,1.8);
	
	% R_A

	\node[draw=none,red!70!green] at (1,-1) {$R_A$};
	\draw[red!70!green!20,line width=.8mm] (0,-.3) -- (.7,-1);
	\fill[red!70!green!20] (-.7,-.4) rectangle (.7,1.8);

	\fill[black!20] (-.4,-.25) rectangle (.4, .25);
	\fill[black!20] (-.4, .45) rectangle (.4, .95);
	\fill[black!20] (-.4,1.15) rectangle (.4,1.65);
	
	\node[draw=none] at (0,0  ) {$101$};
	\node[draw=none] at (0, .7) {$100$};
	\node[draw=none] at (0,1.4) {$001$};
	
	% Arrows
	
	\fill[\mygrey] (.4,0  ) circle (.1);
	\fill[\mygrey] (.4, .7) circle (.1);
	\fill[\mygrey] (.4,1.4) circle (.1);
	
	\draw[\mygrey,line width=.8mm,->] (.4,0  ) -- (2,0  );
	\draw[\mygrey,line width=.8mm,->] (.4, .7) -- (2, .7);
	\draw[\mygrey,line width=.8mm,->] (.4,1.4) -- (2,1.4);
	
	\draw[line width=.4mm,decorate,decoration={brace,amplitude=5pt}] (-.5,-.3) -- (-.5,1.7);
	\draw[line width=.4mm,decorate,decoration={brace,amplitude=5pt,mirror}] (.5,-.3) -- (.5,1.7);
	
	% Q_A

	\node[draw=none,green!70!red] at (5,-1) {$Q_A(101)$};
	\draw[green!70!red!20,line width=.8mm] (3.2,-.3) -- (4.2,-1);
	\fill[green!70!red!20] (2.25,-.325) rectangle (5.55,.325);
	
	\node[draw=none,green!70!red] at (5.5,2.4) {$Q_A(100)$};
	\draw[green!70!red!20,line width=.8mm] (4.5,.7) -- (5.5,2.1);
	\fill[green!70!red!20] (2.25,.375) rectangle (4.55,1.025);
		
	\node[draw=none,green!70!red] at (1,2.4) {$Q_A(001)$};
	\draw[green!70!red!20,line width=.8mm] (2.8,1.7) -- (1.8,2.4);
	\fill[green!70!red!20] (2.25,1.075) rectangle (4.55,1.725);

	% First Q_A
	
	\fill[black!20] (2.5,-.25) rectangle (3.3, .25);
	\fill[black!20] (3.5,-.25) rectangle (4.3, .25);
	\fill[black!20] (4.5,-.25) rectangle (5.3, .25);
		
	\node[draw=none] at (2.9,0  ) {$001$};
	\node[draw=none] at (3.9,0  ) {$100$};
	\node[draw=none] at (4.9,0  ) {$101$};
	
	\draw[line width=.4mm,decorate,decoration={brace,amplitude=3pt}] (2.4,-.3) -- (2.4,.3);
	\draw[line width=.4mm,decorate,decoration={brace,amplitude=3pt,mirror}] (5.4,-.3) -- (5.4,.3);
	
	% Second Q_A
	
	\fill[black!20] (2.5, .45) rectangle (3.3, .95);
	\fill[black!20] (3.5, .45) rectangle (4.3, .95);
	
	\node[draw=none] at (2.9, .7) {$100$};
	\node[draw=none] at (3.9, .7) {$101$};
		
	\draw[line width=.4mm,decorate,decoration={brace,amplitude=3pt}] (2.4,.4) -- (2.4,1);
	\draw[line width=.4mm,decorate,decoration={brace,amplitude=3pt,mirror}] (4.4,.4) -- (4.4,1);

	% First Q_A
	
	\fill[black!20] (2.5,1.15) rectangle (3.3,1.65);
	\fill[black!20] (3.5,1.15) rectangle (4.3,1.65);
		
	\node[draw=none] at (2.9,1.4) {$001$};
	\node[draw=none] at (3.9,1.4) {$101$};
			
	\draw[line width=.4mm,decorate,decoration={brace,amplitude=3pt}] (2.4,1.1) -- (2.4,1.7);
	\draw[line width=.4mm,decorate,decoration={brace,amplitude=3pt,mirror}] (4.4,1.1) -- (4.4,1.7);

\end{tikzpicture}
\caption{\small For this example matrix $A$, any pair of rows $(r, q) \notin P_A$ will never allow a valid extension of $A$. We encode the set of compatible pairs $P_A$ as the set $R_A$ where each entry $r$ relates to a set $Q_A(r)$. In this example, notice that even though $A$ is OR$^*$, it does not satisfy the symmetry rules. }
\label{fig:PRQ}
\end{center}
\end{figure}
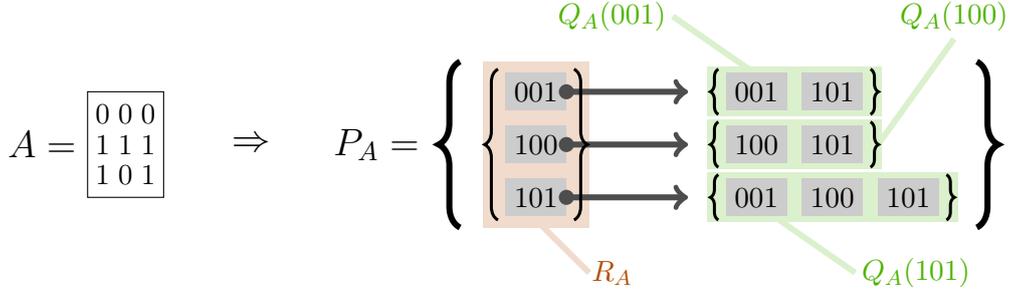

\end{defi}

%Based on the information contained in the stored sets of compatible pairs, we can easily compute the set of possible (non-symmetrical) extensions to $A$:
%\begin{align} \label{eq:possible-extensions}
%	N_A \triangleq \Big\{ r \in R_A : \left[\begin{smallmatrix} A \\ r \end{smallmatrix}\right] \text{ is OR and satisfies the symmetry rules} \Big\}.
%\end{align}
%Here, the symmetry rules impose that the columns of the matrix are lexicographically sorted and that the first and second rows are respectively all zeros and all ones. 
Given an OR$^*$ matrix $A$, the set of possible extension rows $q$ such that $\left[\begin{smallmatrix}A \\ q\end{smallmatrix}\right]$ is OR$^*$ can be easily identified using $P_A$. 
%Moreover, pairs of rows can only become incompatible with the already fixed rows hence the following lemma. 
Moreover, the $P_A$ set can only shrink as we add rows to $A$ hence the following lemma.
\begin{lem} \label{thm:extend-branch}
	Let $A$ be some $d \times n$ Order-Regular$^*$ matrix of the form $\left[\begin{smallmatrix} A^- \\ r \end{smallmatrix}\right]$ and let $A^+ = \left[\begin{smallmatrix} A \\ q \end{smallmatrix}\right]$ for some row $q$. Then $A^+$ is Order-Regular$^*$ iff $q \in Q_{A}(r)$. Furthermore for any $r, q$, it holds that $(r,q) \in P_{A^+}$ iff both $(r,q) \in P_{A^{}}$ and there exists a column $k$ such that $A^+_{d,k} \neq A^+_{d+1,k} = r_k = q_k$. Therefore $P_{A^+} \subseteq P_{A^{}}$.
	\begin{proof}
		First we observe that: 
		\begin{align*}
			q \in Q_A(r) & \Leftrightarrow (r,q) \in P_A, \\
			& \Leftrightarrow \forall\, i, 1 \leq i < d, \exists\, k : \hspace{.7cm} A^{}_{i,k} \neq A^{}_{i+1,k} = r_k = q_k, \\
			& \Leftrightarrow \forall\, i, 1 \leq i < j = d, \exists\, k :  A^+_{i,k} \neq A^+_{i+1,k} = A^+_{j,k} = A^+_{j+1,k},
		\end{align*}
		since $A^{}_{i,k} = A^+_{i,k}$ for all $i, 1 \leq i \leq d$. Furthermore, using the fact that $A$ is OR$^*$, we also have that for all $i, j, 1 \leq i < j < d$, there exists a column $k$ such that $ A^+_{i,k} \neq A^+_{i+1,k} = A^+_{j,k} = A^+_{j+1,k}$. Therefore, condition~\eqref{eq:OR} is verified for all $i,j, 1 \leq i < j < d+1$ and we have that $q \in Q_A(r)$ iff $A^+$ is OR$^*$.
		
		The fact that $(r,q) \in P_{A^+}$ iff both $(r,q) \in P_{A^{}}$ and there exists a column $k$ such that $A^+_{d,k} \neq A^+_{d+1,k} = r_k = q_k$ follows directly from the definitions of $P_{A^{}}$ and $P_{A^+}$.
	\end{proof}
\end{lem}
%To extend a branch corresponding to a matrix $A$, we only have to compute the sets $P_A$ and $N_A$ which can be done efficiently.
%\begin{cor} \label{thm:compute-branch}
%	Let $A$ be some $d \times n$ Order-Regular matrix of the form $\left[\begin{smallmatrix} A^- \\ r \end{smallmatrix}\right]$ and let $A^+ = \left[\begin{smallmatrix} A \\ q \end{smallmatrix}\right]$ for some $q \in N_A$. Then $P_{A^+}$ and $N_{A^+}$ can be computed using respectively $O\left(|P_A|\right)$ and $O\left(|Q_{A^-}(r)| + |R_A|\right)$ OR checks.
%	\begin{proof}
%		From Lemma~\ref{thm:extend-branch}, computing $P_{A^+}$ requires to check one OR constraint for each pair in $P_A$. In the same spirit, computing $N_{A^+}$ requires to check two OR constraints and to make an equality test of two rows for each row in $Q_{A^-}(r) \cap R_A$. Of course, checking an OR constraint can be done in constant time.
%	\end{proof}
%\end{cor}
%Overall, using the sets of compatible rows allows us never having to check the same quadruplet of rows more than once in a given branch, and this at the cost of a reasonable amount of memory.

{\bf Direct cutting.} Storing and maintaining the sets of compatible pairs of rows during the search has an additional advantage. Assume we are looking at a branch corresponding to a $d \times n$ matrix $A$. Then we have $|R_A|$ distinct rows appearing as $r$ in the set of compatible pairs of rows $P_A$. There is clearly no way of getting more than $d+|R_A|+1$ rows by extending this particular branch. Consequently, when searching for an $(m^*+1) \times n$ OR$^*$ matrix, if $d+|R_A|+1 < m^*+1$, then we discard the node right away and make a step back in the search tree. This idea is formalized by the following lemma.
\begin{lem} \label{thm:cutting}
	Let $A$ be some $d \times n$ Order-Regular$^*$ matrix, let $R_A$ be defined by equation~\eqref{eq:r-projection} and let $m^* = d + |R_A|$. Then there exists no Order-Regular$^*$ matrix with more than $m^*+1$ rows such that the first $d$ rows equal $A$. 
	\begin{proof}
		First we observe that: 
		\begin{align*}
			q \in Q_A(r) & \Leftrightarrow (r,q) \in P_A, \\
			& \Leftrightarrow \forall\, i, 1 \leq i < d, \exists\, k : \hspace{.7cm} A^{}_{i,k} \neq A^{}_{i+1,k} = r_k = q_k, \\
			& \Leftrightarrow \forall\, i, 1 \leq i < j = d, \exists\, k :  A^+_{i,k} \neq A^+_{i+1,k} = A^+_{j,k} = A^+_{j+1,k},
		\end{align*}
		since $A^{}_{i,k} = A^+_{i,k}$ for all $i, 1 \leq i \leq d$. Furthermore, using the fact that $A$ is OR$^*$, we also have that for all $i, j, 1 \leq i < j < d$, there exists a column $k$ such that $ A^+_{i,k} \neq A^+_{i+1,k} = A^+_{j,k} = A^+_{j+1,k}$. Therefore, condition~\eqref{eq:OR} is verified for all $i,j, 1 \leq i < j < d+1$ and we have that $q \in Q_A(r)$ iff $A^+$ is OR$^*$.
		
		The fact that $(r,q) \in P_{A^+}$ iff both $(r,q) \in P_{A^{}}$ and there exists a column $k$ such that $A^+_{d,k} \neq A^+_{d+1,k} = r_k = q_k$ follows directly from the definitions of $P_{A^{}}$ and $P_{A^+}$.
	\end{proof}
\end{lem}
Using this trick, we are able to spot poor branches early on and hence to significantly reduce the size of the search tree without missing any OR$^*$ matrix with $34+1$ rows or more.

\subsection{General implementation}

Combining the ideas from Section~\ref{sec:large-matrices-general}, we sketch the \emph{branch search} strategy in Algorithm~\ref{algo:branch-search}.
%\footnote{Note that in this section, we sometimes use standard Matlab notations when convenient.}
Notice that the starting branch needs not necessarily be the empty matrix. Though, choosing a $d \times n$ OR$^*$ matrix $A$ as the root in Algorithm~\ref{algo:branch-search} will result in an OR$^*$ matrix whose $d$ first rows correspond to the rows of $A$. As we will see, this option will be useful later, but then this also means Algorithm~\ref{algo:branch-search} only performs an exhaustive search on a restricted portion of the tree.

\algsetup{indent=.8cm}
\algsetup{linenodelimiter=.}
\begin{algorithm}
\caption{Branch search}
\label{algo:branch-search}
\begin{algorithmic}[1]
	\renewcommand{\algorithmicrequire}{\textbf{Input:}}
	\REQUIRE $A$, the (PS)OR$^*$ matrix of size $d \times n$ at the root of the search tree (optional, $[\,]$ by default).
	\renewcommand{\algorithmicrequire}{\phantom{\textbf{Input:}}}
	\REQUIRE $m^{\mathrm{target}}$, the target number of rows for the solution $A^*$ (optional, $0$ by default).
	\vspace{.2cm}
	\renewcommand{\algorithmicrequire}{\textbf{Initialization:}}
	\REQUIRE Precompute $P_A$ using equation~\eqref{eq:compatible-pairs}.
	\vspace{.2cm}
	\renewcommand{\algorithmicensure}{\textbf{Output:}}
	\ENSURE $A^* =$ $\mathrm{branchsearch}(d, A, P_A, A)$, a (PS)OR$^*$ matrix with $n$ columns and the maximum 
	\renewcommand{\algorithmicensure}{\phantom{\textbf{Output:}}}
	\ENSURE (or the target) number of rows such that the $d$ first rows of $A^*$ are given by $A$.
	\vspace{.2cm}
	
	\renewcommand{\algorithmicloop}{\textbf{function} $\mathrm{branchsearch}(\ell, A^{(\ell)}, P^{(\ell)}, A^*)$}
	\renewcommand{\algorithmicendloop}{\textbf{end function}}

	\LOOP
		\IF{$\ell > \mathrm{\# rows}(A^*)$} 
			\STATE $A^* := A^{(\ell)}$.
		\ENDIF
		\IF{$\mathrm{\# rows}(A^*) = m^{\mathrm{target}}$} 
			\RETURN $A^*$.
		\ENDIF
		\STATE Extract $R^{(\ell)} := R_{A^{(\ell)}}$ from $P^{(\ell)}$ using equation~\eqref{eq:r-projection}. \label{algo-line:bs-R}
		\STATE Extract $Q^{(\ell)} := Q_{A^{(\ell)}}(r)$ from $P^{(\ell)}$ using equation~\eqref{eq:q-projection} with $r$ being the last row of $A^{(\ell)}$. \label{algo-line:bs-Q}
		\IF{$\ell + |R^{(\ell)}| < m^{\mathrm{target}}$ \label{algo-line:bs-cutting}} 
			\RETURN $A^*$.
		\ENDIF
		\FOR{$q \in Q^{(\ell)}$ \label{algo-line:bs-iterate}}
			\STATE $A^{(\ell+1)} := \left[\begin{smallmatrix} A^{(\ell)} \\ q \end{smallmatrix}\right]$.
			\IF{$A^{(\ell+1)}$ satisfies the \emph{symmetry rules}} 
				\STATE Compute $P^{(\ell+1)} := P_{A^{(\ell+1)}}$ using equation~\eqref{eq:compatible-pairs}. \label{algo-line:set-of-compatible-pairs}
				\STATE $A^* :=$ $\mathrm{branchsearch}(\ell+1, A^{(\ell+1)}, P^{(\ell+1)}, A^*)$.
			\ENDIF
		\ENDFOR
		\RETURN $A^*$.
	\ENDLOOP
	
	\vspace{.2cm}
	\renewcommand{\algorithmicrequire}{\emph{Symmetry rules:}}
	\REQUIRE the columns of the matrix must be lexicographically sorted and 
	\renewcommand{\algorithmicrequire}{\phantom{\emph{Symmetry rules:}}}
	\REQUIRE the first and second rows must be respectively all zeros and all ones.
\end{algorithmic}
\end{algorithm}

{\bf Complexity issues.} The steps~\ref{algo-line:bs-R} and~\ref{algo-line:bs-Q} can be performed efficiently using, e.g., a two dimensional array to encode the $P_{A^{(\ell)}}$ sets. Moreover, the step~\ref{algo-line:bs-cutting} encodes the direct cutting according to Lemma~\ref{thm:cutting}. Regarding step~\ref{algo-line:bs-iterate}, the rows $q$ can be taken in any order. By adding randomness in the order, we allow the algorithm to return any matrix with the target size. Finally, Lemma~\ref{thm:extend-branch} ensures that step~\ref{algo-line:set-of-compatible-pairs} requires at most $|P_{A^{(\ell)}}|$ OR$^*$-checks which is still the most expensive operation of each step of the recursion. Observe that since $Q_A(r) \subseteq R_A$ (as shown in the proof of Lemma~\ref{thm:cutting}), it holds that $|P_A| \leq |R_A|^2$ and hence that the cardinality of both sets decrease together when $\ell$ increases.

\subsection{For extremal matrices: we need exhaustive search}

To further speed up our code in order to perform an exhaustive search on all OR$^*$ matrices with 7 columns, we develop a code capable of parallel processing.

{\bf Parallelization.} In Algorithm~\ref{algo:branch-search}, it is possible to perform the search in parallel on different branches of the tree. For this purpose, we first fix some depth $d$ and precompute every possible non-symmetrical $d \times 7$ OR$^*$ matrix. These matrices act as the roots of several independent subtrees that together span the complete search tree. We then launch Algorithm~\ref{algo:branch-search} in parallel each time with a different root matrix as input. It finishes with the answer whenever every subtree has been completely searched. 

In our case, we chose $d = 9$ which resulted in 106 million distinct subtrees of variable size. We obtained $35$ subtrees that ended up with an OR$^*$ matrix of $33+1$ rows but none with an OR$^*$ matrix of $34+1$ rows, leading to the statement of Theorem~\ref{thm:conjecture-fails} in Section~\ref{sec:fibonacci-conjecture}.
%We found an OR matrix with $33$ rows for $35$ of these subtrees but none resulted in a matrix with 34 rows, hence the following theorem.

%\begin{thm} \label{thm:conjecture-fails}
%	For $n = 7$, there exist no Order-Regular matrices with $34 = F_{n+2}$ rows and therefore Conjecture~\ref{thm:hansen-zwick-conjecture} fails.
%\end{thm}
%
%Of course, a formal proof of Theorem~\ref{thm:conjecture-fails} would require a proof of correctness for our code. While we cannot provide such a proof, our implementation of Algorithm~\ref{algo:branch-search} is available on demand. We also encourage anyone interested to implement their own variant of Algorithm~\ref{algo:branch-search} and thereby confirm our result.

\subsection{For building blocks: we need an efficient search strategy}

To search for (P)SOR building blocks with $8$ columns, the strategy described in Section~\ref{sec:large-matrices-general} still applies but the size of the search space does not allow to perform an exhaustive search. However, in this case, we only need to find a large matrix but not to prove that it is the largest (we found an SOR block with 33 rows and a PSOR block with 35 rows in our case, see Figures~\ref{fig:building-block} and~\ref{fig:softened-building-block}). To this end, based on the special structure of (P)SOR matrices, we designed an efficient search strategy to quickly find these large instances. We now develop this strategy for SOR matrices. An equivalent strategy exists for PSOR matrices as well but it requires to introduce some nonessential details. As in Section~\ref{sec:large-matrices-general}, we here use the variation of the (P)SOR condition denoted by (P)SOR$^*$ and defined in Remark~\ref{rem:variation-OR}.

The search strategy is based on the reversing operation that reverses the order of the rows and negates the even rows.
\begin{defi}[Reversing] \label{def:reverse}
	Let $A \in \{0,1\}^{m \times n}$. We define $A^{\mathrm{rev}}$, the \emph{reverse} of $A$, where for all $1 \leq i \leq m$, we have:
	\begin{align*}
		A^{\mathrm{rev}}_{i,k} \ = 
		\begin{cases}
			A_{m+1-i,k} & \text{ if $i$ is odd,} \\
			1 - A_{m+1-i,k} & \text{ if $i$ is even,}
		\end{cases}
	\end{align*}
	for all columns $k$. 
\end{defi}
The key observation is that reversing an SOR$^*$ matrix preserves its Strong Order-Regularity$^*$.
\begin{lem} \label{thm:back-and-forth}
	If $A \in \{0,1\}^{m \times n}$ is SOR$^*$, then its reverse is also SOR$^*$.
	\begin{proof}
		For all $i, j, 1 \leq i < j < m$, let $i' = m-j$ and $j' = m-i$ so that $1 \leq i' < j' < m$. Let also $I(i,j) \triangleq \begin{bmatrix} i & i+1 & j & j+1 \end{bmatrix}$ and $I'(i,j) \triangleq m+1-I = \begin{bmatrix} j'+1 & j' & i'+1 & i' \end{bmatrix}$ using Matlab notations. From the Strong Order-Regularity$^*$ of $A$, there exist two columns $k_1$ and $k_2$ such that:
		\begin{align*}
			A_{I'(i,j),k_1} = \begin{bmatrix} \alpha & \alpha & \alpha & \overline{\alpha} \end{bmatrix} \quad \text{and} \quad A_{I'(i,j),k_2} = \begin{bmatrix} \beta & \beta & \overline{\beta} & \beta \end{bmatrix}
		\end{align*}
		for some $\alpha, \beta \in \{0,1\}$. Then for $A^{\mathrm{rev}}$, the reverse of $A$, we have:
		\begin{align*}
			\begin{cases}
				A^{\mathrm{rev}}_{I(i,j),k_1} = \begin{bmatrix} \alpha & \overline{\alpha} & \alpha & \alpha \end{bmatrix} \quad \text{and} \quad A^{\mathrm{rev}}_{I(i,j),k_2} = \begin{bmatrix} \beta & \overline{\beta} & \overline{\beta} & \overline{\beta} \end{bmatrix} & \text{if $i$ is \emph{odd} and $j$ is \emph{odd},} \vspace{.2cm} \\
				A^{\mathrm{rev}}_{I(i,j),k_1} = \begin{bmatrix} \alpha & \overline{\alpha} & \overline{\alpha} & \overline{\alpha} \end{bmatrix} \quad \text{and} \quad A^{\mathrm{rev}}_{I(i,j),k_2} = \begin{bmatrix} \beta & \overline{\beta} & \beta & \beta \end{bmatrix} & \text{if $i$ is \emph{odd} and $j$ is \emph{even},} \vspace{.2cm} \\
				A^{\mathrm{rev}}_{I(i,j),k_1} = \begin{bmatrix} \overline{\alpha} & \alpha & \alpha & \alpha \end{bmatrix} \quad \text{and} \quad A^{\mathrm{rev}}_{I(i,j),k_2} = \begin{bmatrix} \overline{\beta} & \beta & \overline{\beta} & \overline{\beta} \end{bmatrix} & \text{if $i$ is \emph{even} and $j$ is \emph{odd},} \vspace{.2cm} \\
				A^{\mathrm{rev}}_{I(i,j),k_1} = \begin{bmatrix} \overline{\alpha} & \alpha & \overline{\alpha} & \overline{\alpha} \end{bmatrix} \quad \text{and} \quad A^{\mathrm{rev}}_{I(i,j),k_2} = \begin{bmatrix} \overline{\beta} & \beta & \beta & \beta \end{bmatrix} & \text{if $i$ is \emph{even} and $j$ is \emph{even}.}
			\end{cases}
		\end{align*}
		In every case, the Strong Order-Regularity$^*$ of $A^{\mathrm{rev}}$ is ensured.
	\end{proof}
\end{lem}

Based on Lemma~\ref{thm:back-and-forth}, we can now formulate our \emph{back-and-forth search} strategy to find large SOR$^*$ matrices as described by Algorithm~\ref{algo:back-and-forth}.

%\begin{algorithm}[h] 
%	\DontPrintSemicolon
%	\KwIn{$d$, $T$.}
%	\KwOut{An SOR$^*$ matrix.}
%	\BlankLine
%	\Initialization{$A^{(0)}$, a random SOR$^*$ matrix with $d$ rows obtained from Algorithm~\ref{algo:branch-search} using input $m^{\mathrm{target}} = d$.} 
%	\InitializationXX{$t = 0$.}
%	\BlankLine
%	\While{stopping criterion}{
%		Compute $B^{(t+1)}$ as the result of Algorithm~\ref{algo:branch-search} using input $A = A^{(t)}$.\; \label{algo-line:baf-algo-1}
%		Compute $B^{\mathrm{rev}}$, the reverse of $B^{(t+1)}$.\;
%		$A^{(t+1)} \triangleq \ B^{\mathrm{rev}}_{1:d, :}$, the first $d$ rows of $B^{\mathrm{rev}}$.\; \label{algo-line:baf-d-first-rows}
%		$t \leftarrow t+1$.\;
%	}
%	\Return $B^{(t)}$.\;
%	\BlankLine
%	\nonl \emph{Stopping criterion:} after at least $T$ steps, stop whenever $B^{(t-T+1)}$ and $B^{(t)}$ have the same number of rows (stagnation in the last $T$ steps).\;
%	\caption{Back-and-forth search}
%	\label{algo:back-and-forth}
%\end{algorithm}

\algsetup{indent=.8cm}
\algsetup{linenodelimiter=.}
\begin{algorithm}
\caption{Back-and-forth search}
\label{algo:back-and-forth}
\begin{algorithmic}[1]
	\renewcommand{\algorithmicrequire}{\textbf{Input:}}
	\renewcommand{\algorithmicensure}{\textbf{Output:}}
	\REQUIRE $d$, $T$.
	\ENSURE An SOR$^*$ matrix.
	\vspace{.2cm}
	\renewcommand{\algorithmicrequire}{\textbf{Initialization:}}
	\REQUIRE $A^{(0)}$, a random SOR$^*$ matrix with $d$ rows obtained from Algorithm~\ref{algo:branch-search} using input
	\renewcommand{\algorithmicrequire}{\phantom{\textbf{Initialization:}}}
%	\vspace{-.5cm}
	\REQUIRE $m^{\mathrm{target}} = d$.
	\REQUIRE $t = 0$.
	\vspace{.2cm}
	\WHILE{\emph{stopping criterion}}
		\STATE Compute $B^{(t+1)}$ as the result of Algorithm~\ref{algo:branch-search} using input $A = A^{(t)}$. \label{algo-line:baf-algo-1}
		\STATE Compute $B^{\mathrm{rev}}$, the reverse of $B^{(t+1)}$.
		\STATE $A^{(t+1)} \triangleq \ B^{\mathrm{rev}}_{1:d, :}$, the first $d$ rows of $B^{\mathrm{rev}}$. \label{algo-line:baf-d-first-rows}
		\STATE $t \leftarrow t+1$.
	\ENDWHILE
	\RETURN $B^{(t)}$.
	\vspace{.2cm}
	\renewcommand{\algorithmicrequire}{\emph{Stopping criterion:}}
	\REQUIRE after at least $T$ steps, stop whenever $B^{(t-T+1)}$ and $B^{(t)}$ have the same number
	\renewcommand{\algorithmicrequire}{\phantom{\emph{Stopping criterion:}}}
	\REQUIRE of rows (stagnation in the last $T$ steps).
\end{algorithmic}
\end{algorithm}

In Algorithm~\ref{algo:back-and-forth}, the parameter $d$ is typically chosen so that applying Algorithm~\ref{algo:branch-search} at step~\ref{algo-line:baf-algo-1} finishes in a reasonable time (so $d$ should be large enough to ensure a manageable size of the search trees) while leaving as much room as possible for the optimization process (so $d$ should not be too large either). When looking for SOR$^*$ matrices with 8 columns, we typically used $d = 14$. Also note that in Algorithm~\ref{algo:back-and-forth}, it is important to avoid getting the same $A$ over and over again. We rely on the randomness introduced at step~\ref{algo-line:bs-iterate} of Algorithm~\ref{algo:branch-search} to always get a random instance of the possible $B$ matrices. Interestingly, Algorithm~\ref{algo:back-and-forth} is guaranteed to improve the solution at each iteration, as stated by the following proposition. However, we cannot guarantee that it will find a globally optimal solution. Therefore it may be useful to restart it until finding a matrix with a suitable number of rows.

\begin{prop} \label{thm:monotonicity-back-and-forth}
	In Algorithm~\ref{algo:back-and-forth}, the number of rows of $B^{(t+1)}$ is always at least as large as that of $B^{(t)}$ for all $t \geq 1$.
	\begin{proof}
		At step~\ref{algo-line:baf-algo-1} of Algorithm~\ref{algo:back-and-forth}, applying Algorithm~\ref{algo:branch-search} with $A^{(t)}$ as the root means performing a search in a subtree of the whole tree where the $d$ first rows are fixed. In this subtree, the matrix $B^{\mathrm{rev}}$ computed at the step $t$ is a feasible solution since from Lemma~\ref{thm:back-and-forth}, it is SOR$^*$ and since from step~\ref{algo-line:baf-d-first-rows} its $d$ first rows match those of $A^{(t)}$. Therefore, the best SOR$^*$ matrix $B^{(t+1)}$ that can be found in the subtree must be at least as good (in terms of its number of rows) as $B^{(t)}$.
	\end{proof}
\end{prop}

\section{Conclusions and perspectives} \label{sec:conclusions}

Prior to this work, the three main candidates to be the asymptotic maximum size of OR matrices were the lower bound $\Theta(\sqrt{2}^n)$, Hansen and Zwick's conjecture $\Theta(\phi^n)$ with $\phi$ the golden ratio and the upper bound $\Theta(2^n/n)$. Our results invalidate the first option and leave hope that the second may be overestimated. 
%tend to indicate that reality should rather lie somewhere between the two first options. 
There are chances that the same bound as for Fibonacci Seesaw also applies here ($O(1.61^n)$ for recall).
%Maybe does the same bound as for Fibonacci Seesaw apply here?

Note that in several cases, it is possible to cast an OR matrix back into an AUSO, including when matrices are obtained from Construction~\ref{def:simple-construction}. 
On the other hand, there exist OR matrices that do not correspond to any AUSO.
Yet, if we could prove that a way back exists for the instances generated from Construction~\ref{def:main-construction}$^*$, then our lower bound from Theorem~\ref{thm:main-bound} would also apply to AUSOs. 

Finally, the analysis of PI through combinatorial matrices could apply to other methods based on a similar iterative principle (that is, methods that, at each iteration, choose a subset of outgoing edges and jump to the antipodal vertex in the sub-cube spanned by these edges). It would indeed be interesting to see if the OR condition can be adapted to variants of PI and if our tools can then be successfully applied.

\bibliographystyle{alpha}
\bibliography{biblio}

\end{document}